%% file: paper27.tex
\documentstyle[11pt,epsfig]{article}
\newcommand{\be}{\begin{equation}}
\newcommand{\ee}{\end{equation}}
\newcommand{\beqn}{\begin{eqnarray}}
\newcommand{\eeqn}{\end{eqnarray}}
\newcommand{\kf}{{\bf k}}
\newcommand{\qf}{{\bf q}}
\begin{document}
\begin{titlepage} 
\hspace*{\fill}\parbox[t]{3.8cm}{DESY 96-238\\hep-ph/9612226} 
\vspace*{1cm}
\begin{center}
\vspace{2cm}
{\bf The Space-Time Picture of the Wee Partons}\\
\vspace{0.3cm}
{\bf and} \\
\vspace{0.3cm}
{\bf the AGK Cutting Rules in Perturbative QCD}\\
\vspace{2cm}
{\bf J.Bartels}\footnote{Supported by Bundesministerium f\"ur Forschung und
        Technologie, Bonn, Germany under Contract 05\,6HH93P(5) and
        EEC Program "Human Capital and Mobility" through Network
        "Physics at High Energy Colliders" under Contract
        CHRX-CT93-0357 (DG12 COMA).}\\
II Institut f\"{u}r Theoretische Physik, Universit\"{a}t Hamburg,\\
D-22761 Hamburg, Germany.\\
\vspace{1cm}
{\bf M.G.Ryskin}\footnote{This is work supported in part by the
Grant INTAS-93-0079 and by the Volkswagen Stiftung.}\\
St.Petersburg Nuclear Physics Institute\\
188350, Gatchina, S.Petersburg, Russia
\end{center}
\vspace{2cm}
{\bf Abstract:} We discuss the space-time picture of scattering amplitudes
with single and double ladder exchange in perturbative QCD. Particular
emphasis is given to the Abramovsky Gribov Kanchelli (AGK) rules which
describe the relative contributions of the diffractive dissociation and  
processes with other multiplicities to the elastic scattering amplitude.
Inside the Pomeron Pomeron interaction vertices we find a new matrix which
describes the transition from one s-cut to another and which goes beyond the
original the AGK rules. \\
\end{titlepage}

\noindent
\section{Introduction}
The role of the space-time structure of reggeon exchanges in
developing a physical picture of high energy hadron scattering has been
recognized many years ago\cite{GIP}. The most important result was given
in a paper by Abramovski, Gribov, Kancheli (AGK) \cite{AGK}, where the
so-called ``cutting rules`` both for unenhanced reggeon diagrams and for
the triple pomeron vertex were derived. These rules state the ratios
between the different types of reggeon cuts, i.e. between the
processes with different multiplicities (secondary particle densities)
in the high energy interactions. In the present paper we will discuss
how these ``old ideas`` are realized in perturbative QCD.\\ \\
Recent interest in this question comes from deep inelastic scattering
at small x and, in particular, from the observation of events with a
large rapidity gap between the diffractively excited state of the
virtual photon and the outgoing proton. It is commonly felt that the
latter process is most easily described in the proton rest frame, and it seems
useful to develop an intuitive picture of the space-time evolution of
this very interesting reaction. Even if the Pomeron in this process is,
at best, only partly perturbative and has a strong nonperturbative component,
we nevertheless expect that the basic space time structure may be 
correctly predicted by the perturbative analysis. In any case,
the observation of diffraction dissociation of the photon at HERA 
indicates that unitarity corrections (double ladder exchange) 
to the standard one-ladder description of $F_2$ are present: 
the AGK rules tell us how such a diffractive state contributes to the 
double ladder exchange in the elastic process $\gamma^* + p \to \gamma^* + p$, i.e. they provide a connection between this particular final state and the
first unitarity corrections to $F_2$. On the other hand, from these rules 
it is not immediately obvious how to combine the space-time picture of
the diffractive dissociation process with that of the elastic process
$\gamma^* + p \to \gamma^* + p$. A careful analysis of perturbative QCD 
allows to answer these questions.\\ \\
Throughout this paper it will be crucial to distinguish between scattering
amplitudes and their imaginary parts (energy discontinuities).
In particular, the space-time structure of a scattering amplitude
and the question for intermediate states are, although closely connected,
not exactly the same. Namely, a space-time analysis 
of a perturbative ${\it scattering\,\, amplitude}$
proceeds as follows: rather than using the usual (covariant) Feynman
rules for evaluating diagrams, one decomposes the Feynman amplitudes
into pieces with different time orderings (``old fashioned perturbation 
theory``). At high energies it can then be shown that the leading contribution
arises only from a specific subset of the different time
orderings, and in this way a natural space time picture of the scattering
process emerges. If, on the other hand, one wants to know which
${\it intermediate\,\,physical\,\,states}$ correspond to this scattering 
process,
one has to analyse, because of unitarity, the ${\it imaginary\,\,
part}$ (more general: the energy discontinuity) of the scattering
amplitude. For individual Feynman diagrams, one very often finds cancellations
between different contributions to the energy discontinuity; a famous 
example is the AFS cancellation ~\cite{AFS,M} in planar two ladder exchange 
diagrams which we will have to briefly recapitulate. As a result of these 
cancellations, the scattering amplitude may look somewhat simpler and may 
lead to the (false) impression that certain intermediate states do not
exist. For reggeon diagrams, the AGK rules state in a simple way which
intermediate states exist and what relative weight they have in an elastic
scattering amplitude. It is possible to give a space-time picture also for 
these imaginary parts of a scattering amplitude, but it is somewhat less 
intuitive. In this paper we will illustrate all these aspects for 
QCD reggeon diagrams with two or four reggeized gluons in the 
t-channel.
As a specific example, we have in mind the process 
$\gamma^* + p \to \gamma^* + p$ and the diffractive dissociation of the 
photon. \\ \\
The outline of this paper is the following.
In sect.2 we briefly review the space-time picture of a simple
ladder exchange in the elastic two-body scattering amplitude, as it has
been derived many years ago in the leading
logarithmic approximation of the scalar $\lambda \phi^3$ field theory.
Although, at first sight, the situation in QCD may seem to be infinitely
more complicated, the leading logarithmic aproximation, the BFKL
Pomeron ~\cite{BFKL}, is in complete agreement with this old and simple 
picture.
This is mainly due to the fact that all Feynman diagrams that contribute
to this approximation can be cast into the ladder structure with
${\it effective}$ vertices and kernels (rungs), and for these
elements the region of integration is analogous to the scalar ladders.
For our present purposes we therefore feel that we may skip details
of the BFKL calculations and simply describe the results in terms of the
simple scalar ladder structure. The main new aspect, which is
introduced by QCD and has not been discussed before, is the reggeization
of the gluon. Due to the color degrees of freedom we have to consider
both positive and negative signature of the exchange channel. All this
discussion of the elastic $2 \to 2$ scattering process can easily be
generalized to the process of the diffractive dissociation.\\ \\
In the following section 3 we discuss the space-time structure of the 
double ladder exchange scattering amplitude and, more general, reggeon 
diagrams with up 
to four reggeons in the t-channel, very much in the same spirit as the 
single ladder. When iterating 
the two-ladder exchange in the t-channel we find the new phenomenon of 
the ``change of the arrows`` of the flow of momenta across the diagrams.
With this new element we are able to give a complete space-time picture of 
QCD diagrams with two and four gluons in the t-channel. \\ \\
In sections 4 and 5 we turn to the analysis of intermediate states of these 
QCD diagrams. We begin with a brief review of the AGK rules, applied to the 
two ladder
exchange. In sections 5 we illustrate the AGK rules for the QCD case.
We perform an analysis of the imaginary parts of diagrams with two and 
four gluons in the t-channel, and we explain how the AGK rules are satisfied.
Particular attention is, again, given to the interaction between two ladders:
we obtain ratios of the different cuts across the interaction vertex.
These relations are new and have never been discussed in ``old`` reggeon
theory.

\section{Space-time structure of the leading diagrams}
\subsection{Elastic Scattering of a Virtual Photon}
In the leading log approximation the high energy behavior is dominated by
ladder diagrams (throughout this paper the term ``leading log`` will
always refer to the total energy of the process under consideration). The 
simplest example (for a scalar theory) is shown in Fig.1a. What we
have in mind is the elastic scattering of a virtual photon with momentum
$q^{\mu}$ off a hadron target (momentum $p^{\mu}$). Strictly speaking, the 
cleanest laboratory for investigating the high energy limit of perturbative 
QCD would be the elastic scattering of two virtual photons. On the other hand,
we expect that the space-time structure as well as the AGK-rules will not be 
altered if we replace the target virtual photon by a proton: throughout this
paper we will assume that this lower part of our scattering amplitude, 
although perturbative QCD is not applicable, will have the same general 
features as obtained from perturbative QCD. The coupling of the gluon ladder 
to the virtual photon is through a quark loop. Since at high energies
this quark loop never produces any logarithm of the energy, it appears as
an energy independent ``form factor`` with some dependence on the transverse
momenta of the gluon legs. For most of our discussion it will be sufficient to
replace the quark loop by a parton line. Throughout this paper 
energy $s=Q^2/x_B$ will be assumed to 
be large, i.e. $x_B$ very small. In order to illustrate the space time
structure we will make a further step of simplification: instead of 
QCD diagrams with quarks and gluons we will begin with (massive) scalar 
diagrams. \\ \\
We will use Sudakov (light-cone) variables
\beqn
q_{i,\mu}=\alpha_i q'_\mu +\beta_i p'_\mu +q
_{i,t,\mu};\;\;d^4q_i=\frac 12
sd\alpha_i d\beta_i d^2q_{i,t};\;\;s=2p'q' \nonumber 
\eeqn
\beqn
q^2=-Q^2;\;\;q'=q+x_B p',\;\;p'\approx p;\;\; q'^2=p'^2=0
\eeqn
(note that, in this notation, $q_t^2 < 0$; transverse momentum is defined
as $\qf^2=-q_t^2$).
The  typical structure of the leading logarithmic longitudinal integration
leads to the following ordering of the $\alpha_i$ in the interval from
0 to 1:
\beqn
0<<\alpha_n<<...<<\alpha_2<<\alpha_1<<1.
\eeqn
Due to the momentum conservation we have approximately 
$\alpha_{(q)i} \approx \alpha_{(k)i} = \alpha_i$.
Then in each $\beta_i$ cell one can 
close the contour of the integration in the lower half plane, where the 
only pole comes from the propagator of the rung 
$1/(q^2_i-m^2+i\epsilon)=1/(s\alpha_i\beta_i+q_{it}^2-m^2+i\epsilon)$
\footnote{For the $\beta_i$-loop, this pole is the only one in the lower half
plane; this is because of the positive sign of $\alpha_i$. The other three 
propagators of this loop have the opposite time direction (i.e. the opposite
$\alpha$ direction), and their poles are in the upper half plane.}, 
and one obtains $\beta_i=(m^2-q_{ti}^2-i\epsilon)/s\alpha_i$.
As a result of the $\alpha_i$-ordering (2),
also the $\beta_i$ are ordered, but in the opposite direction:
\beqn
0<<\beta_1<<...<<\beta_{n-1}<<\beta_n<<1.
\eeqn
From (3) it follows that $\beta_{(q)i} \approx\beta_{(k)i+1}=\beta_i$. The 
last s-channel pole 
(the rung at the bottom of Fig.1a) $1/(q^2_n-m^2)$ is used for 
the $\alpha_n$ integration (because of the symmetry between the $\alpha_i$ and
the $\beta_i$, we also could have started with (3), then performed the 
$\alpha_i$ integrals and ended with the ordering 
prescription (2). We then would be left with the $\beta_1$-integral and the
rung at the top of Fig.1a). For an amplitude with the positive signature 
(say, pomeron exchange) we have to sum
up the graphs Fig.1a and 1b. In this sum the leading logs in the real part -
$\int d\alpha_n (\frac{1}{\alpha_n s+i\epsilon}+
\frac{1}{-\alpha_n s+i \epsilon})$ - cancels, and the
main contribution comes from the imaginary part of the propagator $\frac
{1}{\alpha_ns+q^2_{tn}-m^2+i\epsilon}=-i\pi\delta (\alpha_ns+q^2_{nt}-m^2)$.
Thus all the s-channel particles are on mass shell. We finally note that
if one chooses $\alpha_1>1$, or violates the ordering eq.(2), all poles
of one (or more) of the $\beta_i$ variables would be on the same side of the 
integration contour, and the integral would vanish.\\ \\
Let us now consider the same ladder diagrams in ``old-fashioned`` 
(noncovariant) perturbation theory in the target rest frame
($p_{\mu} =(m_N,0,0,0)$, $q_{\mu}'=(Q^2/2x_B m_N,0,0,Q^2/2x_B m_N)$. ``Time`` 
runs from the left to right,
and we draw the vertices in a time-ordered way. In the high energy limit only 
the time ordering shown in Fig.2a survives. For each intermediate state
we draw a cutting line (Fig.2a), and we have to integrate over the time
difference $\tau=t_i-t_{i-1}$:
\beqn
\frac 1{2E_i}\int^{\infty}_0e^{i(E_{cut,i}-q_0+i\epsilon)\tau}d\tau
\eeqn
Here $E_{cut,i}$ stands for the sum of the energies of all lines that belong 
to the intermediate state, $q_0$ for the energy of the incoming photon,
and the life-time corresponding to 
the cut state is proportional to $1/(E_{cut,i}-q_0)$.
In the high energy approximation the energy of a parton with momentum $k_i$
is given by $E_i\sim k_{iz}+\frac{\kf^2_{i}+m^2}{2E_i}$. Due to the momentum
conservation in $E_{cut,i}-q_0$ the 3-components drop out, and in the 
remainder the line with the smallest energy dominates. Making use of 
$E_i \approx \alpha_i q_0'$ and the ordering condition (2) one sees that
the life time of the intermediate state in Fig.2a is proportional to
the energy of the parton with momentum $k_i$, i.e. 
$\tau_i \sim \alpha_i Q^2/2x_B (\kf_{i}^2+m^2)m_N$. For the 
the central cut which ``runs through the target`` we have the energy 
difference $\sum E_i - E_{q+p}$: from the analysis above we know that all 
produced particles are on-mass shell, and $\sum E_i = E_{p+q}$. Hence there
is no limit on the life time of the central cut, i.e. this state can live
for a long time. For the parton life times
before the interaction we have the ordering condition: 
\beqn
\tau_n << \tau_{n-1} << ... << \tau_1
\eeqn
(similarly for the life times after the interactions),
i.e. the parton closest to the incident photon has the longest life-time.
\\ \\
It is instructive to interpret the same Feynman diagrams also in
another reference frame, e.g. the Breit frame (``photon rest frame``) where
the four momentum of the proton has the form $q^{\mu}=(0,0,0,\sqrt{Q^2})$.
This is the frame where the partonic interpretation of deep inelasic scattering
is most easily obtained.
It is now the partons close to the fast proton which have the longest
life-time, since  in this case the energy of the $i$-th parton is mainly
given by the $\beta_i$-component of the momentum $q_i$, which according to (2)
increase with $i$. Therefore
in this frame $\tau_i \approx 2E_i /(\kf_i^2 + m^2) \propto \beta_i$,
and the parton with the longest lifetime is the one at the bottom of the
diagram. Consequently, in this frame the proton splits into partons which
have smaller and smaller longitidinal momenta $\beta_i$, and the photon finally
interacts with the parton for which $\beta=x_B$.
\\ \\
Needless to say that the physics of the process does not
change if we perform a simple Lorentz boost: in each frame
we encounter the feature that the fastest parton has the longest
lifetime. In the proton rest system we probe the partons
with small momentum fraction $\alpha$ inside the photon, in the Breit
system those with small $\beta$ inside the proton. If we
choose a frame in between, e.g. the CM system, it is the interaction
of the small-$\alpha$ partons from the photon with the small-$\beta$
partons from the proton (see Fig.2b). \\ \\
It should be stressed that, up to now, the discussion of the space-time
structure holds for the scattering amplitude and not for the imaginary part.
This follows from the observation that the direction of time is
determined by the relative sign of $\alpha$ and $\imath \epsilon$. For
the elastic amplitude (Fig.1a) the space time picture looks as Fig.2a.
In order to get a precise information on what kind of intermediate
state is contained in our amplitude, we have to ask for cross sections
which, via the unitarity equations, are related to the energy
discontinuity. The space-time picture then becomes quite different:
if, for example, we cut the ladders as indicated in Fig.3a, the time to
the right of the cutting line goes in the
backward  (inverse) direction. This part of the diagram now corresponds
to the complex conjugate amplitude $A^*$ and, hence, it has the
opposite sign
of the $i\epsilon$'s in the $1/k_i^{'2}$ propagators. In other words, the
cut diagram Fig.3a reflects the space-time structure of the $2\to n+1$ matrix
element squared. To say this in different words: taking
the discontinuity of the scattering amplitude is like ``interupting
the time evolution at a certain time`` and then reversing the
direction of time (Fig.3b)
. This point was discussed in more detail in
the appendix of ref.\cite{G2}. \\ \\
Since QCD contains, in addition to the even signature Pomeron exchange,
also the reggeizing gluon with odd signature, a few words have to be
said about the graphs which describe the exchange of a reggeized
gluon. Because of the famous bootstrap property (for further discussion see
section 5), the exchange of a single reggeized gluon is identical to 
ladder diagrams in the color octet representation.
The simplest diagram of such a type is, again, a ladder of the type Fig.1a,b.
But now, in contrast to the even signature (Pomeron) case, where we
take the sum of the ladder and its crossed counterpart (Fig.1a and
Fig.1b) now we deal with the difference $(\frac 1{\alpha_n s}-\frac
1{-\alpha_n s})$. This simple difference leads to a change in the space-time 
structure. Namely now the $\alpha_n$ integral from the lowest rung 
gives a logarithm which is not cancelled by the crossed diagram, and it
comes from a region of integration where the virtuality of this rung is large,
i.e. in contrast to the even signature case there is now one rung off its mass 
shell. In order to obtain the correct interpretation of this effect we have   
again to study the space-time picture (Fig.2a). The major change compared to 
the even signature state is results from the off-shellness of the lowest rung:
now we have no longer the equality $\sum E_i=E_{p+q}$, and the time integral 
belonging to the central cut leads to finite lifetime 
$\Delta \tau \sim 1/(\sum E_i - E_{p+q}) \approx \alpha_n s$.
So the life time of the full odd signature ladder is substantially smaller
than that of an even signature ladder. 
This difference becomes particularly accute if we consider a reggeized
gluon inside a larger diagram (Fig.4). Although
inside this ladder we have the usual ordering of life-times, compared to
its ``environment`` the ladder represents only a short-lived fluctuation.
Therefore, as far as the leading real part of this subamplitude is concerned
one can consider this ladder as a single object --  as a
reggeized gluon. However, if we consider the (nonleading) imaginary part of 
the reggeon, which corresponds to the discontinuity 
of the propagator $i\pi\delta (m^2-q_n^2)$ at the bottom of the ladder 
one looses one energy logarithm and has again the normal time-life 
$\tau_i\sim E_i/(m^2+\kf^2_{i})$.\\ \\
So far we have used scalar field theory to illustrate the space time structure
of Feynman diagrams in the high energy limit. As we said in the introduction,
this structure holds also for the QCD scattering amplitude if we simply 
replace the line of the incoming particle by a quark-antiquark pair, the 
ladder rungs by BFKL kernels, and t-channel scalar particles by reggeized 
gluons. \\ \\
\subsection{Diffractive Dissociation}
Now let us consider the scattering amplitude of the diffractive dissociation 
of the beam particle.
To be definite, we consider the dissociation of the virtual photon into
a system with fixed mass $M$ ($M^2 \ll s$ which consists of a $q\bar{q}$ pair 
plus a few gluons (Fig.5a). 
As before, we begin with the rest frame of the target proton.
The corresponding picture is shown in Fig.5b.                                 
The largest time is the formation time of the particles in the diffracted
system $M^2$; the quarks have the largest values of $\alpha_i\sim O(1)$. 
The gluon emitted from the quark has already a smaller $\alpha$, but it is 
still larger than that of gluon with momentum $k$ etc. The time
interval occupied by the t-channel pomeron is controlled by
the life-time of the gluon with momentum $k$; its energy is the same as
that of the rung with $q$. Hence
$\tau_k\propto \alpha q_0/\qf^2$. It is even better to express the time
$\tau$ in terms of the $\beta_i$ variables, which denote the momentum fraction
of the proton momentum $p_{\mu}$ carried away by the parton-i.
As $\beta_{i}=\qf^2_{i}/\alpha_is$, we have ~\cite{I}
$\tau_i\sim 1/\beta_im_N=1/x_im_N$, where $m_N$ is the nucleon (target) mass.
It should be stressed that while at the bottom of the graph (Fig.5b) the time
and the $z$ longitudinal intervals are small (of the order of proton radius;
as the corresponding $x_i$ here are large enough ($\sim 0.3-0.1$) and
$\tau\simeq 1/x_im_N\sim 1Fm$), the gluon at the top of the ladder needs a 
rather long time
 $\sim 1/x_P m_N$ ($x_P\ll 1$) to recollect all the partons which have been
emitted inside the Pomeron ladder.\\  \\
The picture looks quite different in the Breit frame where the proton 
is fast and the beam ``is at rest`` (Fig.5c). The first impression 
is that now the partons produced inside the Pomeron belong to the fast proton.
Nevertheless, it is exactly the same Feynman graph as in Fig.5a. Our different 
ways of re-drawing the same Feynman diagrams only illustrate that any 
interpretation involving ``time`` is not Lorentz invariant and hence changes
from the one reference frame to another.
In the Breit frame $\tau\propto \beta_i$, and the time interval at
the bottom of the pomeron ladder is the largest one, but even
at the upper end of the pomeron the life-time of the highest parton with 
momentum fraction $x_1$ is larger than the formation time for the particles 
in the diffracted system $M^2$. However these particles will be detected far 
away from the interaction point and they do have time enough to absorb 
(interact with the) our 'non-local' pomeron.\\ \\

\section{Diagrams with Two or Four t-Channel Gluons}
Having demonstrated that amplitudes for both the elastic process 
$\gamma^*+p\to \gamma^*+p$ and the diffractive process 
$\gamma^*+p \to M+p$ have a simple
space time interpretation, it seems most natural to ask how the latter
process contributes to the total cross section $\gamma^*+p$ which is
given by the imaginary part (or the energy discontinuity) of the elastic 
scattering amplitude for $\gamma^*+p$. This leads us to a study of 
higher order corrections (Fig.6) to the elastic process, diagrams with two 
or four gluons in the t-channel, and we will have to analyse the imaginary
part of this amplitude. For the moment, however, we will postpone this
question, and in this section we shall consider the space-time structure
of the scattering amplitude of this larger class of diagrams.\\ \\
Counting first powers of $\alpha_s$ and $\ln s$ and
comparing with the single ladder, leading $\ln s$ approximation, we
have lost two powers of $\ln s$. A consistent QCD analysis
requires to study {\it all} contributions which contribute to this level 
of accuracy, i.e. we are not allowed to simply limit ourselves to diagrams 
with two-ladder exchange but have to include much more general 
classes. Generally speaking, it is expected that they can be divided into two
groups. The first one preserves the one-ladder structure but introduces
corrections to the BFKL kernel; we expect that will have the same space-time
structure as the BFKL ladders itself. The second group of corrections, for 
which Fig.6a presents a simple example, goes beyond the one-ladder structure
and contains both the t-channel iterations of the two-ladder states and 
the enhanced
diagrams (single-ladder state between the coupling to the external
particle and the two-ladder states). An example is shown in Fig.6b. These 
contributions are expected to be of particular relevance for the unitarization
of the BFKL Pomeron. In this 
paper we will restrict ourselves to this latter class of corrections. As we 
shall see they contain new elements of the space time picture.
\\ \\
As to the four gluon state, it is important to note that the
interactions between the four gluons can be arranged in two different ways.
Either we can group the pairwise interactions into ladders (e.g.
between the lines (12) and (34)). We then have to sum over all ``switches``
from one coupling scheme to another (e.g. from (12)(34) to (13)(24)) 
and over all iterations in the t-channel.Any such
switch can also be viewed as an interaction between two ladders.
Alternatively, we could reorder the
rungs in the four gluon state in such way
that we obtain a geometric series of the kernel $\sum_{ij} K_{ij}$
where the sum extends over all pairs (ij) of gluon lines. It is easy to see
that both ways of summing over all rungs lead to the same result. For most
of our discussion it will be more convenient to adopt the first method of
summation, i.e. to speak about ``ladders` and their ``interactions``.\\ \\
\subsection{Two-Ladder Exchange}
Let us organize our analysis in several steps. For simplicity, we first 
return to scalar field theory. We begin with a two ladder
diagram and investigate the coupling of the ladders to the incoming particle. 
We will show that
the leading log contributions comes only from those graphs where one of the 
reggeons can be put inside the other one (Fig.7a,maximal nested structure) 
~\cite{CHSS}. The diagrams that we are going to analyse are illustrated in 
Fig.7a and b. As before, in order to get the maximum number of the logarithms 
one has to order the 'sizes' of the reggeons. 
Let us compare the diagrams Fig.7a and 7b. For the graph
7a it is enough to choose the $\alpha '\gg\alpha "$ and there will be
no any problem with the integrations over the $\beta$-variables. One
can close the contour of the $\beta_R$ (the momentum fraction
transfered through the reggeon as a whole) on the pole $1/q^2_1$, the
$\beta "$ contour - as usual, on the pole $1/q^2_2$ and then the $\beta
'$ - on the pole $1/q^2_3$. In the case of the graph Fig.7b the situation
 is quite different. After the $\beta_R$ integration here we get the
two poles ($q_2$ and $q_3$) with the largest
$\alpha_2\simeq\alpha_3\simeq1$ in the lower half plane of the $\beta
"$ integral. The leading contributions coming from this poles cancel
 each other\footnote{As the two poles with the largest $\alpha$ are
enough to provide the convergence of the $\beta$
integral, one can omit the $\beta$ in all the other propagators
(with the smaller value of $\alpha_i$) and then closing the contour of
integration in the upper half plane (where there are no any
singularities now) get the zero.}
and finally the graph 7b does not gives the leading logs. From the
 space-time point of view it is evident. The life-time of the partons
$\tau '\gg\tau "$ (at $\alpha '\gg\alpha "$)  does not satisfy the
required time ordering condition. The parton $\alpha "$ in Fig.7b should be
aborbed after the parton $\alpha '$, but its life-time is too small to
provide this. Thus the configuration shown in Fig.7b does not
contribute.\\ \\
If we choose the wrong ordering $\alpha '\ll\alpha "$
(i.e. the time-live $\tau '\ll\tau "$ we lose the leading log
also in Fig.7a. Now for the $\beta_R$-integration the two leading poles 
(with largest values of the $\alpha$-variables) will both lie in the lower
half plane of the $\beta_R $ plane.
Thus we have show that the time interval occupied by the upper parton
of the external reggeon $\tau '\propto \alpha '$ should be much larger
than the value of $\tau "\propto \alpha "$, which corresponds to the
inner reggeon. This properties of the two reggeon cut kinematics comes
originally from the famous S.Mandestam paper\cite{M} and was discussed
with respect to QCD in \cite{GLR}.\\ \\
It should be noted that only the life-time of the top level partons 
in each reggeon ladder of Fig.7a have to be strongly ordered. 
About the other partons which are not so close to the
vertex of the two reggeon emission we can only say that the values of $\alpha_i'$
and $\alpha_i"$  are ordered in each reggeon separetely
$$0<<\alpha_n'<<...<<\alpha_2'<<\alpha_1'<<1.$$ $$0<<\alpha_n"<<...<<
\alpha_2"<<\alpha_1"<<1.$$
and there are no any special
relations between the time-life (i.e. the values of $\alpha_i'$ and
$\alpha_i"$) of the partons from different reggeon.\\ \\
At the lower end of the diagram Fig.7a the picture is exactly the
same. The best way to see it is to go to the beam particle ($Q$) rest
frame. In this frame the lowest partons have the largest time-life
$\tau_{n-1}\propto\beta_n$, and we have to put $\beta_n'\gg\beta_n"$ or
$\beta_n"\gg\beta_n'$. As it was mentioned above the ordering at the
bottom does not correlate with the ordering at the top of the graph and
any combination of the regions $\beta_n'\gg \beta_n"$ or
$\beta_n"\gg\beta_n'$ and $\alpha '\gg\alpha "$ or $\alpha "\gg\alpha
'$ do gives the leading log contribution. Of course each
time one has to draw the Feynman diagram in such a way
that the reggeon with the
largest $\alpha$ would be the external one at the top and - one with the
largest $\beta$ - at the bottom of the diagram. \\ \\
The generalization to QCD is straightforward. The discussion given in this
section applies directly to the 
case where both gluon ladders couple to the same quark line. But since we 
have to sum over all possibilities of attaching the ladders to the quark 
antiquark pair, we also have to say a few words about the other configurations.
As a rule we state that, in order to contribute to the leading logarithmic
approximation, the life time of the first rung of one ladder has to be larger
than that of the second ladder. In particular, this implies that in the 
contribution where one ladder couples to the quark, the other to the antiquark,
only those time orderings survive where one ladder ``is fully contained``
inside the other.\\ \\
\subsection{The Space Time Picture of a Two-Pomeron Interaction}
As the next element of our general class of diagrams illustrated in Fig.6b
we analyse the
space-time structure of the interaction between two ladders (Fig.7c - f).
For simplicity we ignore the complications due to the reggeization of the
gluon and the complex structure BFKL rung, and consider the scalar case.
Figs.7c,d, and f are examples of switches from one way of pairing to another.
For example, starting in Fig.7d from below, we first have the pairing
(12)(34), then, after one rung between line 1 and 3, the pairing (14)(23).
Equivalently, one might say that the ladders (12) and (34) interact with
each other and then continue as ladders (14)(23).
The switch from the lower pair of ladders to the upper ones can,
alternatively, be viewed as the interaction between two ladders, i.e.
as a $2 \to2$ Pomeron vertex. Cutting Fig.7d by a vertical line between the
second and the third t-channel line we obtain intermediate states with
double multiplicities in the upper half. The main result to be derived in
the following is that, for diagrams of this type, the second and third
t-channel lines must have {\em changes in the direction of time}. 
(the observation that partons may have negative fractions of momentum
has also been mentioned in ~\cite{BKL}). To illustrate this
statement, let us remind that in two noninteracting ladders (14) and (23)  
(Fig.7a) along lines 1 and 2 the time arrows always point downwards,
along llines 3 and 4 upwards. Obviously, rungs always connect t-channel
lines with opposite time directions. The new feature to be explained in the
following is that, as soon as we try to introduce rungs between lines 1 and 2
(or 3 and 4), the time flow along lines 2 and 3 must change its direction.
\\ \\
Before we discuss this feature of Fig.7d in more detail, let us first show that
not all switches lead to a nonzero contribution, even if we allow for changes
in the time direction. For example, one might try
to draw a configuration with zero density in the central region (Fig.7c). 
In the beginning, let us consider the situation where we have the
usual ordering along the second t-channel line (i.e. the line in Fig.7c 
which carries $\alpha_2$ should be drawn with the arrow in the opposite 
direction). Then one sees easily that the t-channel gluons 1 and 2 (or 3 and
4) cannot interact with each other. Indeed, in order to receive
the leading log in the $\alpha_{i}$ integration one has to close the 
$\beta_i$ contour around the pole $1/q^2_i=1/(\alpha_i\beta_is-\qf^2_{i})$ with
the largest $\alpha$ in the whole $\beta_i$ loop (see Fig.1a). This pole 
gives us $1/\alpha_is$ and then the integral $\int^1_{\kf^2/s}
\frac{sd\alpha_i}{s\alpha_i}\propto lns$.
The contributions of the $\alpha_j\beta_is$ ($j>i$) terms in all other 
propagators, say,  ($1/k^2_{i+1}\,$,$\,1/k'^2_{i+1}\,,...$) 
are small in this case and we may  neglect the
$\alpha_{i+1}\beta_is=(\qf^2_{i}+m^2)  \frac{\alpha_{i+1}}{\alpha_i}$ 
term in comparison with $q^2_{i,t}$, because of the $\alpha$ ordering (2)
($\alpha_{i+1}<<\alpha_i$).
So, if the value of $\alpha '$ (dashed line in Fig.7c) is smaller than
$\alpha_3$ we lose the log in the $\alpha_3$ loop. The term $\alpha_3\beta 's$
becomes the dominant one in the propagator $1/k^2_3$, and the integral over
$\alpha_3$ looks as $\int d\alpha_3/\alpha^2_3$). If, on the other hand
we change in Fig.7c the direction of the line $\alpha_2$ (as drawn in
Fig.7c), we still get zero. Namely, if $\alpha
'>>\alpha_3$ then $\alpha_2\simeq -\alpha '$ becomes negative. For 
the $\beta_2$ integration we then have two poles in the lower half plane,
and they cancel each other
\footnote{This two poles with the largest
$\alpha\;\;(\alpha =\alpha_1$ or $\alpha =-\alpha '$ provide the convergence
of the $\beta_2$ integration in the region $|\beta_2|\sim \qf^2_t/\alpha 's$.
Therefore we can omit the $\beta_2$ terms in other propagators (where it
contribution is small) and then close the contour in the upper half plane where
there are no any singularities now. Thus the $\beta_2$ integration gives the
zero result.}.\\ \\
Now let us return to Fig.7d and show that by changing the arrows
for the two vertical gluon lines we obtain a nonvanishiung contribution.
As before the 
arrows indicate the sign of the $\alpha_i$ variable (i.e. the direction of
time) and the vertical position of each horizontal line reflects the $\alpha$
ordering ($\alpha_1>>\alpha_2>>\alpha_3>>...$).
The dashed lines show the structure of the $\beta_i$ loops, while the doubled
(dashed) part of this lines mark the propagators with the largest $\alpha$ in
each $\beta$ loop. Points, where the arrow changes its direction (sign of
$\alpha$) are marked by open circles.
Almost all the integrations in the graph Fig.7d look as usual. For example, the
$\beta_4$ integration picks up the pole $1/q_4^2$, the $\beta_5$ integration 
the pole $1/q^2_5$, $\beta_0$ the pole $1/q^2_0$ and so on. Two exceptions 
are the $\beta_2$ and $\beta_3$ loops, where pairs of poles ($1/q^2_2$
and $1/k'^2_3$ for $\beta_2$, and $1/q^2_3$ and $1/k^2_5$ for $\beta_3$) lie
in the lower complex halfplane. To be sure that these pairs of  poles 
do not cancel we have to choose $\alpha_{(k)3}\approx \alpha_3 >>
\alpha_{(k')3} \approx \alpha_4$ and 
$\alpha_4>>\alpha_5$. The convergence of the $\beta_2$ and $\beta_3$ 
integrations are garanteed by the poles $1/q^2_2$, $1/k^2_3$ and 
$1/q^2_3$, $1/k^2_4$, resp., which are located in opposite half planes.
The contributions of $\beta_2$ and $\beta_3$ terms in all other propagators 
are negligible.\\ \\
The general condition which, for example in Fig.7d, leads to the nonvanishing 
contribution
is the following: below the vertices marked by a circle on line 2 and 3
we have to change the signs of the $\alpha$ variables, i.e. the direction of
time. Namely, the $\alpha$ ordering has to fulfill the requirement that, 
for a given $\beta$-loop, the $\alpha$-value on the vertical line opposite
to the marked vertex is much larger than in the vertical line below the
marked vertex: $\alpha_3 \gg \alpha_3'$. This is the formal way to check 
whether a given time ordering
will lead to a nonzero contribution. \\ \\
A more intuitive way to distinguish
between vanishing and contributing time orderings is the following:
in order to have a nonvanishing contribution it must be
possible to draw the space time diagram (see, for example, Fig.7e) in such
a way that one of the two ladders is fully inside the other. In other words,
emission (and absorption) of the first rung of the outer ladder must happen
before (and after) the emission (and absorption) of the rung of the inner
ladder. In Fig.7e we show the space time picture of Fig.7d: obviously we can
move the marked vertices until one ladder is totally inside the other.
In Fig.7f we present another example, with the crossing
symmetry (with respect to the gluons 2 and 3) amplitude.\\ \\
\subsection{The Full Diagram}
Let us finally collect our results and attempt to give a space-time 
description of the scattering amplitude of a general diagram Fig.6a or b.. 
For simplicity we restrict ourselves to the 
reference frame where the photon is fast and the proton is at rest.
Beginning at the upper end where the photon dissociates into the quark-
antiquark pair, we consider the case where two gluons are radiated from 
(and later on reabsorbed by) the quark pair. What we have shown is that, 
depending on the coupling of the four gluon lines to the quark pair,
only selected time orderings are relevant, For example, if all gluon
couple to the same quark line only those configurations survive
where the first s-channel gluon produced from the first t-channel gluon 
has to be 
reabsorbed last, and the decay products of the second gluon have to 
recollected first. In other words, the ``inner rungs`` must be fully 
contained inside the ``outer rung``. This restriction holds only for the 
first rung in each
ladder. After this, the formation of ladders between lines 1 and 4, and lines
2 and 3 may continue: each reggeon then continues its own cascade in such a 
way that the life times of the rungs is getting shorter and shorter. As a part
of this requirement, a rung can connect only lines with opposite time
direction. After the first rung in each ladder, there is 
no further correlation in time between the rungs of the first and the second
reggeon.\\ \\ 
After a few steps it may happen that the two ladders interact, e.g. a rung 
from 1 goes to 3 or from 2 to 4. This requires no change in the time direction.
However, if a rung emitted from 1 wants to be absorbed by 2, line 2 has to 
change its direction. The example of Fig.7c demonstrates that even with this
change of the time direction we may get zero. As stated at the end of 
section 3.2, in order to have a nonzero contribution we have to be able to
draw the space time structure in such a way that the new ladders (12) and (34)
can be put one inside the other. The formal way is, again, to check that
the leading poles of the $\beta$ integrals (i.e. the ones with largest
$\alpha$-values) always lie on different sides of the real axis.
Fig.8 shows two more examples, illustrating the space time structure of 
more complex configurations.
Alternatively, we can also have the situation that initially only one gluon 
is emitted from the quark pair (and later absorbed again). In this case,
below the quark pair first a single reggeon ladder develops, as described 
in the beginning of this paper. After some time, one gluon rung emits and 
absorbs an extra gluon: this four gluon system now develops in the same way 
as described in the previous paragraph. In particular, initially the two
ladders starting at the gluon rung have to be maximally nested again.\\ \\
Finally there are less symmetric configurations: at the top the one of the two
quarks produces a gluon, and the rungs connected to this gluon may end at two
different lines. In this way two ladders develop which have one line
in common. After a while, there may be a rung between line 2 and 3: in order
that such a rung has the correct time structure, line 2 has to change its
time direction.\\ \\
\subsection{A Remark on the Diffractive Dissociation}
It may be worthwhile to explore in more detail why certain time orderings
do not contribute to the high energy behaviour of the scattering amplitude. 
In many cases, the formal way of eliminating a particular time ordering
is the cancellation of poles in $\beta$-integrals. This is nothing else
but saying that different contributions to the imaginary part of the
scattering amplitude cancel against each other. In the following, we 
illustrate this cancellation for a case of particular interest, the AFS 
diagram shown in
Fig.9a and b. The reason why this configuration plays a special role is the
following. From the discussion of the previous section it follows that
this configuration does not contribute to the scattering amplitude (it is 
not maximal-nested). 
On the other hand, the contribution to the imaginary part 
which follows from putting the two-particle intermediate state on mass-shell
(Fig.9c), represents the square of ``elastic scattering ``. So, at
first sight it seems mysterious how the (very important) elastic 
cross section (Fig.9c) disappears inside the scattering amplitude
of Fig.9b. The resolution of this somewhat puzzling situation follows from
the old arguments of Amati, Fubini, Stangellini ~\cite{AFS} and Mandelstam
~\cite{M}. \\ \\
Let us first review the argument for the simple scalar $\lambda \phi^3$
theory; later on we shall demonstrate that QCD works the same way.
Beginning with the space time picture in Fig.9b it is quite obvious why, in
the scattering amplitude, this
configuration cannot be realized: on the one hand, the longitudinal distance 
between the partons n' and n'' should be larger than the life time 
$\tau_1 \sim E_1/\qf_t^2$ of the rungs at the top. On the other hand, at high
energies this separation will be much larger than the size of the target which
characterizes the separation between the points p' and p'' at the bottom of
Fig.9b. Next, looking into the high energy formula of this diagram one finds
that it vanishes since for the central $\beta$-integral
the three leading poles (with the largest $\alpha_i$) lie all on the same 
side of 
the integration contour: closing the contour in the opposite half plane gives
zero. Equivalently, the sum of the three poles adds up to zero. This 
cancellation can also be seen if we decompose the imaginary part
of the diagram. Three contributions belonging to low multiplicity final 
states are illustrated in Fig.9c and d (not shown is the conjugate of
Fig.9d), and they can be shown to cancel each 
other. The first contribution (Fig.9c) is positive, and it describes the 
elastic cross section. The two other terms (Fig.9d and its conjugate) are 
negative, since the virtuality $q_0^2$ is positive while the propagator 
$1/k''^2$ in Fig.9d is negative.\\ \\
In this way, in $\lambda \phi^3$ theory, the planar diagram of Fig.9a does 
not contribute to the elastic scattering amplitude. Nevertheless,  
elastic and diffractive dissociation intermediate state exist and give 
a nonzero contributions. Their sum, however, adds up to zero.
In QCD we have the same pattern of cancellations, but - mainly because 
of the vector nature of the gluon - some important details
are different. First we have to remember that, when we draw reggeon diagrams
with BFKL interaction kernels, we are, in fact, summing over larger classes of
Feynam diagrams and combining them into reggeons and ``effective`` nonlocal 
vertices. For
example, in a convenient axial gauge $p'_{\mu}A^{\mu}$
the BFKL effective gluon production vertex is obtained from the sum of
three Feynman diagrams shown in Fig.10a. Therefore, in order to demonstrate 
the AFS cancellation
in Fig.9a (now with reggeized gluon lines in the t-channel, and BFKL 
vertices for the rungs), we need to decompose the BFKL kernels on both sides.
This is illustrated in Fig.10b where we decompose the upper rung inside the
left hand ladder. In Fig.10c we draw the corresponding discontinuities which,
according to the AFS mechanism, are needed in order to cancel the elastic 
intermediate states. \\ \\
In order to show in more detail that in QCD this 
cancellation really works in the same way as in the scalar theory, a few
comments are in place. First, since the triple gluon vertex $V_3$ in the first
diagram Fig.10b has
a nontrivial dependence upon the virtualities of its external lines, we
have to check that Fig.10c really has the same
expression. Working in the axial gauge $p_{\mu}'A^{\mu}$ the gluon
propagator takes the form
\beqn
\frac{d_{\mu \nu}(q^2)}{q^2},\;\; 
d_{\mu \nu}(q^2)=g_{\mu \nu}-\frac{q_{\mu} p_{\nu}'+p_{\mu}' q_{\nu}}
                 {qp'}
\eeqn
with $d_{\mu \nu} p'^{\mu}=0$ and $d_{\mu \nu}q^{\nu} \simeq -\frac{q_{t\mu}}
{\alpha_q}$. As it is well-known ~\cite{FGL}, in the high energy limit a 
t-channel gluon is dominated by the longitudinal ``nonsense`` polarization, 
i.e. $\sum_{\lambda} e_{\mu}^{\lambda} e_{\nu}^{\lambda} 
 \approx p_{\mu}'q_{\nu}'/(q'p')$. In our gauge it means
that at the upper end the polarization vector $e_{\mu}$
of the gluon is $p_{\mu}'$,
whereas at the lower end one contracts with $q_{\nu}'$: 
$d_{\mu \nu}(k)q'^{\mu} = -\frac{2k_{t\nu}} {\alpha_ks}$. Now the elementary
vertices shown in Fig.11 take the form: $\Gamma_a=-2k_{t\sigma}$,
$\Gamma_b=\alpha_i \delta_{\sigma \sigma'}$, $\Gamma_c=-\frac{2q_{t\sigma}}
{\alpha_q}$, and with the help of $k^2=-\kf_t^2$, $q_s^2=\qf_t^2/\alpha_q$
it is easy to see that the three graphs in Fig.10b sum up to 
the effective BFKL vertex:
\beqn
-\frac{2k_{t\sigma}}{k^2} - \frac{2 q_{t\sigma}}{\alpha_{q}q_s^2}
= \frac{2}{\kf^2 \qf^2} (\qf^2 k_{t\sigma} - \kf^2 q_{t \sigma}).
\eeqn 
Applying the same expressions for the vertices (with $q_s^2 =\qf^2/\alpha_q$) 
to the first graph in Fig.10c, we see that in fact it reproduces the first 
term of the BFKL vertex, while the second and the third graphs in Fig.10c
reproduce the last term of eq.(7).\\ \\
Another point to be mentioned is that the left hand part of the second 
diagram in Fig.10b, as a 
Feynam diagram, does not contribute to the leading logarithmic approximation.
The two contributions coming from the $\beta$ poles of the $1/q_s^2$ and
the $1/q_s'^2$ propagators cancel each other (i.e. the AFS cancellation
repeats itself inside a single ladder). This cancellation 
reflects the wrong space time structure of this diagram: the life time of the
gluon with momentum $q"$ is larger than the separation between the points 2 and
1. On the other hand, when computing the BFKL ladder with a fixed number
of gluons in the final state, the pole of the $q_s$ line is taken into
account in order to produce the BFKL vertex on the rhs of the ladder, whereas
the pole of the $q_s'$ line goes into the reggeization of the upper left 
t-channel gluon.\\ \\
Both the AFS cancellation and the cancellation of poles in the second
diagram of Fig.10b are examples illustrating that, at high energies,
a Feynman amplitude may give a zero contribution to the scattering amplitude,
but at the same time may contain nonzero contributions to partial cross
section. Other examples of this phenomenon can be found in ~\cite{BW}.\\ \\
\section{The AGK Cutting Rules for the Exchange of Two Regge Poles}
So far we have tried to develop a space-time picture of the scattering
amplitude for the process $\gamma^* + p \to \gamma^* + p$, generalizing from
the one-ladder approximation to diagrams with up to four gluons in the
t-channel (in the s-channel, as we have demonstrated, this corresponds to 
higher order partonic interactions). Now we turn to the question of
intermediate states, in particular the contribution of the diffractive
dissociation. As we have said before, this question has to be 
answered by an analysis of the contributions to the imaginary part (or
energy discontinuity), i.e. by a study of the unitarity content of our 
diagrams. Here we will make use of the AGK cutting rules ~\cite{AGK}. \\ \\
It may be helpfull to first briefly review the content of these rules. 
For a two-reggeon exchange (Fig.12) the scattering amplitude is of the form
~\cite{G1}:
\beqn
T=-is \int d\Omega_2
N (i\xi_{\alpha_1}s^{\alpha_1 -1}) (i\xi_{\alpha_2}s^{\alpha_2-1}) N
\eeqn 
Here the $N$'s denote the (real-valued) partial waves above and below the 
two reggeons in Fig.9
(which by themselves may contain Regge cut singularities), $\alpha_1$ and
$\alpha_2$ are the reggeon trajectory functions, and the signature 
factors are
\beqn
\xi_{\alpha} = \frac{e^{i\pi \alpha} \pm 1}{\sin \pi \alpha}
\eeqn
The detailed form of the phase space integral in eq.(8) is not of importance
for our present dicussion and can be found in ~\cite{AGK}; in fact, we 
only need the phase structure of (8):
\beqn
T \sim -i (i\xi_1) (i\xi_2)
\eeqn
From (8) we find
\beqn
2 Im T = 2\left (Re \xi_1 \,Re \xi_2 - Im \xi_1\, Im \xi_2 \right).
\eeqn
Now the AGK rules state that the same result can be obtained by summing 
seperate contributions to the
imaginary part. Namely, the diffractive cut $\sigma_0$ with zero
multiplicity  in the central region (Fig.12, the cut runs between the two 
reggeons), 
the multiperipheral cut $\sigma_1$ with single multiplicity 
(cutting one of the two reggeons), and the double multiperipheral 
cut $\sigma_2$ with double multiplicity (cutting simultaneously 
both reggeons) give
\beqn
\sigma_0 & = & 2\left (Re \xi_1 \,Re \xi_2 + Im \xi_1\, Im \xi_2 \right),
            \nonumber \\
\sigma_1 & = & - 8 Im \xi_1 \, Im \xi_2     \nonumber \\
\sigma_2 & = & 4 Im \xi_1 \, Im \xi_2,
\eeqn
resp. (the factors 2, 4, and 8 count the different possibilities 
of interchanging reggeon 1 and 2). Taking the sum of these cuts,
$\Sigma \sigma = \sigma_0 + \sigma_1 + \sigma_2$, we obtain, in fact, 
the result (11). In ~\cite{AGK} one also finds a generalization to an
arbitrary number of reggeons.\\ \\
The most interesting case is the two Pomeron
exchange (even signature with intercept close to one). Here we can
neglect the real part of the signature factors, and for the ratios of the 
various cuts we obtain:
\begin{equation}
\sigma_0\,:\,\sigma_1\,:\,\sigma_2\,:\,\\ \Sigma \sigma\;=\;1\,:\,-4\,:\,2\,:\,-1
\end{equation}
\\ \\
To clarify the physical meaning of the relation (13) we will discuss the
scattering of a hadron h on a deuteron target.
In the impulse approximation the cross section is simply 
$\sigma^{(0)}(hd)=\sigma (hp)+\sigma (hn)$. 
However, for some part of the time the impact parameters 
$b_t$ of the proton and neutron inside the deuteron may coincide and they 
shade each other. Let us 
assume, for simplicity, that the nucleons inside the deuteron are the grey 
disks with constant opacity $\xi <1$ ($\xi =1$ corresponds to a black disk) 
and area $S$, and put $b_{t,p}=b_{t,n}=0$. Then the inelastic cross section 
for the scattering of the hadron h on a single nucleon is simply $S\xi$.
In this reaction, the final state has single density of the secondary hadrons. 
Due to the optical theorem it is easy to obtain the elastic cross. The 
imaginary part of the forward scattering amplitude is given by:
\beqn
2Imf_{el}(b)\;=\;|f_{inel}(b)|^2\,+\,|f_{el}(b)|^2
\eeqn
Assuming that $\xi\ll1$, and making use of the fact that at high energies
the scattering amplitude with positive signature amplitude is imaginary,
we get $\sigma_{el}(hN)=S(\xi /2)^2$. Thus the sum of cross sections is:
\begin{equation}
\sigma (hp)\,+\,\sigma (hn)\;=\;2(\sigma_{inel}\,+\,\sigma_{el})\;=\;2S\xi\,+\,S\xi^2/2
\end{equation}
This is the impulse approximatiom result. \\ \\
The corrections to this result come from different sources: \\
i) near the second target(nucleon) the initial flow of the beam hadrons
has become smaller by a factor $1-\xi$,  due to the absorbtion on the 
first target. This gives a correction $\Delta\sigma_{1(i)}=-S\xi^2$.\\
ii) after the interaction with the first target with the probability $\xi$ 
the next inelastic collision with the second nucleon takes place, leading
to the doubled multiplicity of the secondaries instead of the single one.
This leads to a further decrease of the cross section $\sigma_1$ by 
$\Delta \sigma_{1(ii)} = -S \xi^2$. Both corrections affect the final states
with single multiplicity. Their sum is 
$\Delta\sigma_1=\Delta\sigma_{1(i)}+\Delta\sigma_{1(ii)}=-2S\xi^2$.\\ 
iii) At the same time we obtain for the double multiplicity final state
$\Delta\sigma_2=+S\xi^2$.\\
iv) Finally, for the elastic scattering cross section on a target with the two
overlapping grey disks one should use for the inelastic cross section in (14)
$2S\xi$ which leads to $\sigma_{el} = S\xi^2$, i.e. a correction to (15) of 
the form $\Delta\sigma_{el}=\Delta\sigma_0=S\xi^2/2$.
Putting all these corrections together we obtain
\beqn
\Delta\sigma_0\,:\,\Delta\sigma_1\,:\,\Delta\sigma_2\;=\;1\,:\,-4\,:\,2
\eeqn
quite in agreement with the AGK rules\cite{AGK}.\\ \\
The physical interpretation of these corrections is, indeed, the same as
suggested by the AGK rules. Namely, starting with the correction 
$\Delta \sigma_2$, which belongs to the final states with double 
multiplicity, it correspond to cutting both ladders simultaneously 
($\sigma_2$ in (13)). The space-time structure is illustrated in Fig.13a. 
The two contributions to $\Delta \sigma_1$ have single
density final states, i.e. they cut only one of the two ladders ($\sigma_1$
in (13)) (Fig.13b). They describe the screening corrections to the single 
multiplicity contribution, due to the rescattering of the hadron h on the
``second`` target. Finally, the correction $\Delta \sigma_0$ corresponds to 
Fig.13c;
it corresponds to the diffractive cut ($\sigma_0$) in (13)). 
It should be stressed that we detect the final hadrons far away
and long time after from the collision. Therefore it is possible to follow the
rescattering in the single amplitude in Fig.13b or c: there is enough time
to create and than to recollect the whole ladder
which describe the reggeon exchange before the secondary hadrons
reach our detector. \\ \\
Finally let us consider a two reggeon exchange amplitude in which the 
reggeons have odd signature with intercept close to one. This case will become 
important when, in the following section, we have to deal with the reggeizing
gluon in QCD. In contrast to the even signture case, 
the signature factor now is approximately real, and from (12) one sees that
the diffractive cut $\sigma_0$ dominates. So the AGK rules lead to a quite
different result. Instead of (13), we now have
\beqn
\sigma_0\,:\,\sigma_1\,:\,\sigma_2\,:\,\Sigma \sigma = 1\,:\,0\,:\,0\,:\,1
\eeqn
In other words, cutting an odd signture reggeon gives a subdominant
contribution compared to the uncut reggeon. In QCD, this means that
cutting a reggeizing gluon brings us down by a power of $\alpha_s$ (i.e.
we are losing one power of $\ln s$), and it is clear that the leading-log
approximation, the BFKL Pomeron ladder, is entirely due to the 
cut which runs in the middle of the ladder and never touches  a reggeized
gluon. It should, however, be noted that the ``old`` AGK rules (13) remain
valid also for the odd signature reggeons, if we restrict ourselves to
the (nonleading) imaginary parts.\\ \\
\section{The AGK rules in QCD}
Now we are ready to discuss the cutting rules for the diagrams with four
t-channel gluons in QCD (left half of Fig.14a).
The simplest way to confront these QCD diagrams with the AGK rules
is the following. We first redraw the sum of all diagrams as reggeon
diagrams with reggeized gluon propagators and BFKL kernels. This amounts
to combining sums of Feynman diagrams in a special way. Next we
consider the t-discontinuity across the four reggeon state.
Loosely speaking, this is equivalent to selecting a four gluon t-channel state
in the center of the diagrams and then summing independently above and below
this state over all t-channel iterations. In other words, instead of
the scattering amplitude itself we consider its t-channel discontinuity across
the four - reggeon cut (Fig.14a).
What appears above and below the selected four reggeon state are
- almost- the t-channel partial waves with four gluons in the t-channel. 
Pairs of two reggeized gluons form bound states, in particular Pomerons with 
vacuum quantum numbers, and we then will be able to compare Fig.14a with 
Fig.12 and  eq.(8). Our goal is the analysis of the different contributions
to the discontinuity in s, i.e. we will have to analyse how the different
s-cut lines pass the selected four gluon state. \\ \\
From our previous discussions it follows that in our analysis of the four
reggeon intermediate state we cannot avoid to discuss also the odd signature
reggeon. Namely, as we have argued in section 3, in the diagrams of Fig.14a 
we are now down by two 
powers of $\alpha_s$ compared to the leading one-ladder approximation, and 
at this level we expect not only to see the two-Pomeron cut but also 
contributions with two cut odd signature reggeons, e.g. the double 
multiperipheral cut of the BFKL Pomeron  (cf. the discussion at the
end of the previous section). Consequently it is to be expected that
the diagrams in Fig.14a contain 
contributions other than the double exchange of two even-signature
reggeon ladders to which the AGK rules (13) apply. The odd-signature 
contributions should be removed (and
considered seperately) before we can compare with the AGK rules.
In order to isolate these contributions we have to
concentrate on those configurations where the two ladders are in an
octet state, i.e. we have to adress the bootstrap problem which we have
mentioned already before.\\ \\
The simplest example for the bootstrap mechanism in QCD is given by the
BFKL ladder diagrams. They represent the leading logarithmic
approximation for the imaginary part of the elastic scattering
amplitude. For the color singlet exchange case this provides the
leading approximation for the amplitude. For the octet exchange,
however, the amplitude which is given by the exchange of a single
reggeized gluon, is real-valued, and the imaginary part is down
by one power of $\alpha_s$. In order to match the leading real part,
the imaginary-valued BFKL ladders, when projected onto the color octet state,
must coincide with the single reggeized gluon exchange. This is, in
fact, the case: the ``sum of the ladders is identical with the single reggeized
gluon``. This identity is referred to as the bootstrap property.\\ \\
For amplitudes with more than two gluons in the t-channel one meets
similar identities. For example, from the LLA analysis given in ~\cite{BW} 
on obtains for the four gluon amplitude $N_4$ in Fig.14a above the 
four-reggeon 
state that, whenever two neighbouring gluons are in the antisymmetric 
color octet state with odd signature, the two gluons ``collaps`` into a single
reggeized gluon. These contributions are then simple BFKL ladders with a 
splitting of the reggeized gluons into two (or even three) gluons at the
lower end. 
Applying this result to the amplitudes in Fig.14a
above and below the four gluon state, one easily sees that, for
example, there will be a contribution where lines 1 and 2 and lines 3 and 4
``collapse`` into single odd-signature gluons. 
Obviously, for these contributions the AGK rules have to be interpreted 
differently from the even signature (as we have discussed in the previous 
section). In ~\cite{BW} a complete analysis of the
four gluon amplitude has been performed, and it has been shown that the
four gluon amplitude which consists of the diagrams illustrated in
Fig.14a can be written as a sum (Fig.14b) of reggeizing pieces and an
``irreducible`` remainder (Fig.14c) which is free from any odd-signature 
admixtures.
In the following we will first consider this latter part.
At the end of this section we shall return to the reggeizing contributions.
\\ \\
After having isolated and removed all those states where ladders
``collaps`` into
reggeized gluons, the remaining ``irreducible`` piece has a very special 
structure.
First of all, it contains only ``enhanced`` diagrams (Fig.14c). This means that
in this irreducible part the four-gluon state does not couple directly to the
external photon. For it has been shown in ~\cite{BW} that all diagrams where 
three or four gluons couple directly to the quark loop belong to the reggeizing
contributions of Fig.14b. What remains for the irreducuble piece Fig.14c are
BFKL ladders between the quark loop and 
the transition vertex: two-gluons $\to$ four gluons (this is a new
element in the BFKL dynamics. It has been derived and discussed in 
~\cite{BW,BLW}). Below this vertex
the four gluon state starts. An important property of the transition vertex 
is its symmetry (both in momentum and color) under the exchange of any two of
the four outgoing gluons. The subsequent two-body
interactions between the four gluons preserve this symmetry, and consequently 
the four gluon state remains completely symmetric. Finally, as discussed at 
the end of section 3, we re-group the interactions inside the four reggeon
state into ladders - e.g. first all pairwise interactions inside the subsystems
(12) and (34), then for the pairing (13), (24) and so on. Because of the
symmetry, the ladders have always even signature, and the amplitude is 
completely symmetry if we interchange any pair of lines. It is important to
keep in mind that the ladders are not restricted to the color 
singlet configuration, i.e. they can be also in the symmetric 8 or 
27-representation.
\\ \\
Putting now together these irreducible pieces from above and from below,
it is easy to compare the different ways of cutting these diagrams.
First we note that a vertical cut across the diagram in Fig.14c does not
alter its (absolute) value: when the line intersect a horizontal
(s-channel) gluon, the discontinuity $2\pi \delta(q_i^2)$ is the same
as the pole contribution (we have argued before that all s-channel
gluons are on the mass shell). This statement applies to both the
rungs inside the four gluon state and the 2 $\to$ 4 transition vertex.
In particular, the transition vertex does not depend upon if and where
we draw a cutting line.
On the other hand, if we intersect a
vertical gluon line we lose one (or more) power of $\ln s$.
Therefore, in the following all cuts will run between t-channel gluon lines. 
Let us pick, inside the four gluon system, a two-ladder state, and let
the discontinuity line run
between the gluon lines ``2`` and ``3`` (Fig.15a). If the rungs above and 
below belong to the
pairing (12)(34), we have the diffractive cut which comes with the
weight $+(\frac{1}{2!})^2$ (this includes the statistics factors
for the two-gluon t-channel states on both sides of the cutting
line). If the rungs belong to the
assignment (13)(24) or (14)(23), we obtain the double multiperipheral 
cut (Fig.15c) with weight $+2(\frac{1}{2!})^2$. Next let the discontinuity line
pass between lines ``1`` and ``2`` (Fig.15b). Now all pairings belong to the
single multiperipheral cut, and together they give the weight $-3\frac{1}{3!}$
(the simplest way to understand the negative sign is by looking at a 
situatiuon where inside the system (234) two of the gluons, say 2 and 3, 
form an even signature Pomeron state. In this case the Pomeron exchange
can be viewed as an absorptive correction to gluon 4, and therefore comes with
a relative minus sign).
The same contribution is obtained from the discontinuity line between the
lines ``3`` and ``4``. In summary, we have the relative weights
\beqn
\sigma_0:\sigma_1:\sigma_2 =  +1/4:-1:+1/2,
\eeqn
which agrees with the AGK cutting rules in eq.(13). \\ \\
Note that in this argument we did not refer to any color coefficients: all 
that went into our argument is the complete symmetry of the four gluon state 
under the exchange of color and momenta of gluon lines. We clearly are 
allowed to conclude that the AGK rules are satisfied for the two-reggeon states
with vacuum color quantum numbers, i.e. Pomerons. But our
analysis also shows that Pomeron ladders are not enough: 
even-signature ladders with nonzero color are present and participate in 
the AGK rules (for these ladders, the cut between the ladders should no longer
be called ``diffractive``). \\ \\
The presence of these color nonsinglet ladders has a very important practical 
consequence. Since the full set (Fig.14a) of
four-reggeon corrections to the scattering amplitude contains also these
non-vacuum quantum number two reggeon ladders, a measurement of
the diffractive dissociation cannot be used to estimate the size of
Fig.14a , i.e. the full unitarity correction. In the diffractive dissociation
one measures, at least for sufficiently large rapidity gaps, only the 
vacuum quantum number ladders on both sides of the cutting line. 
The diagrams in Fig.14a, on the other hand, also contain color nonzero ladders:
their final states are expected to have similar characteristics as the
BFKL ladders. In particular, the produced s-channel gluons are 
color-connected, and the probability for events with a rapidity gap is 
small.\\ \\
So far we have concentrated on the two-ladder (four-reggeon) intermediate 
state in the center of Fig.14a; this was the original form of the AGK
relations.
Next one could try to trace the discontinuity line above (or below) the
four gluon state that we have considered so far. On other words, we
now consider Fig.14c and study how a cutting line changes from, e.g., 
``diffractive`` to ``multiperipheral``. Rearranging, inside the four gluon 
state of Fig.14c, the rungs in terms of ladders, we immediatley see that at any
switch (i.e. a change of the pairing (ij)(kl) to (ik)(jl)) the
nature of a cutting line may change. For example, at the switch
(12)(34) $\to$ (13)(24) the discontinuity line between line ``2`` and
``3`` changes from ``diffractive`` to `` multiperipheral``. On the other
hand, the line between ``1`` and ``2`` remains ``multiperipheral``.
Considering all possibilities, we arrive at the following
transition matrix:
\beqn 
\begin{array}{|r||c||c||c|} \hline
from: & \sigma_0 & \sigma_1 & \sigma_2 \\ \hline \hline
to: \sigma_0 & 0 & 0 & 1/2 \\ \hline
\sigma_1 & 0 & 1 & 0 \\ \hline
\sigma_2 & 1 & 0 & 1/2 \\ \hline
\end{array}
\eeqn
Let us note that this matrix (19) is nontrivial: instead of the usual
(1:-4:2) relation we now find that there are no transitions from single
to double (or zero) densities cuts, only from single to single, from
double to double, double to zero, and from zero to double multiplicity.
Furthermore, this matrix conserves the initial AGK
ratios (13). Indeed, if at the beginning of the four gluon
state we had the relations
$\sigma_0\,:\,\sigma_1\,:\,\sigma_2\;=\;1\,:\,-4\,:\,2\;$ then after
the $2\to 2$ pomeron interaction (which includes
the sum of all type of the vertices
(14)(23) $\leftrightarrow$ (12)(34), (14)(23) $\leftrightarrow$
(13)(24), (12)(34) $\leftrightarrow$ (13)(24)\footnote{The
contributions of each of them are equal to each other\cite{B1}} )
we get $2\cdot\frac 12: -4\cdot 1:(1\cdot1+2\cdot\frac 12)=1:-4:2$
again. In other words, the vector $(1,-4,2)^T$ is an eigenvector to
the matrix (19).Thus in any rapidity interval the relative contributions
of the processes with zero : single : doubled $\;$ densities of the
secondary particles remain 1:-4:2.\\ \\
Nevertheless, taking into accout the pomeron-pomeron interactions we get
different two-particle rapidity correlations. As it was shown
in \cite{AGK}
and was discussed in more detail in \cite{LR}
the admixture of the two pomerom cut contribution to the single pomeron
(ladder) exchange leads to the nonzero positive correlation in the inclusive
two particle distribution:
\beqn
R_2=\frac{d^2N}{dY_1dY_2}/(\frac{dN}{dY_1}\frac{dN}{dY_2})-1>0
\eeqn
(here $N$ is the number of particles at the given rapidity $Y$).
This correlation is the long range one. The value of $R_2$ (due to the pomeron
cut with $\alpha_P(0)=1$) is constant (i.e. does not dependent on 
$\Delta Y=Y_1-Y_2$ in the whole rapidity range occupied by the pomeron cut).
Now after the ladder-ladder interactions the correlation $R_2(\Delta Y)$
becomes smaller as with the probability equal to 1/2 (see table(19)) the double
multiperipheral cut turns into the  diffractive one. The value of $R_2$ is
still positive, but  falls down with $\Delta Y$. The characteristic correlation
length $\Delta$ ($R_2$ becomes small at $|\Delta Y|<\Delta$) reflects the
average distance between the $2\to 2$ pomeron vertices.\\ \\
Let us finally return to the reggeizing contributions (Fig.14b) which we 
have left out so far. Starting point is the decomposition illustrated in 
Fig.14b, and we have to insert these reggeizing pieces above (and below) 
the four gluon state in Fig.14a. The sum in the second line of Fig.14b 
includes color octet odd signature reggeons (i.e. the reggeized gluon), 
but also color octet
even signature reggeons as well as color singlet even signature reggeons.
Moreover, the reggeons may split into two or three gluon propagators.
Finally, we have to sum over permutations of the lower lines. 
The complete formula can be found in ~\cite{BW}. For our discussion
we need to mention only one of its most important properties, namely the 
symmetry
under the exchange of gluon lines. For the even-signature reggeons splitting 
into two gluons, the sum of all reggeizing terms is completely symmetric 
under the simultaneous exchange of momenta and color indices. In an obvious 
notation, the sum can be written as (12)(34)+(13)(24)+(14)(23) 
where the labels refer to both color and momentum
assignment. Consequently, we have a complete symmetry under the exchange of
gluon lines. For the 
odd-signature reggeons (the reggeized gluon), we have a mixed symmetry: we
again sum over all pairings: i.e. (12)(34)+(13)(24)+(14)(23), but for each 
pairing
(ij)(kl) we now have antisymmetry under the exchange of i and j or k and l.
For the terms where one of the two reggeons splits into three gluons, it 
carries the gluon quantum numbers, and we have the sum
1(234)+2(134)+(124)3+(123)4. So we have symmetry under the exchange of any two
gluons. For each term i(jkl), however, the subsystem (jkl) has a more complex
structure. Either the pair (jk) or the pair (kl) are in a state with definite 
color and signature, such that the system (jkl) forms an odd signature
color octet.\\ \\
When inserting these reggeizing pieces into Fig.14a, we generally have three
contributions: reggeizing pieces from above and below, reggeizing pieces only
above or only below the four-reggeon state. The two latter contributions 
again  
describe a transition from the two-reggeon state to a four reggeon state;
because of the symmetry properties, the AGK rules work the same way as
discussed above (note that the odd signature reggeons drop out: since the
four gluon state from below is completely symmetric, it cannot couple to
an odd signature reggeon). So all we have to discuss is the first combination, the reggeizing terms from above and below.\\ \\ 
The sum of all these terms, once more, splits into several categories.
First we have the ``diagonal terms`` (Fig.16a), where the reggeons from 
above and below match. These terms have the structure of eq.(8), and it is 
straightforward to discuss the AGK rules. For the even-signature reggeons
from above and below the counting is exactly the same as for the two-reggeon
ladders above: we simply replace ``cut ladder`` by ``cut reggeon`` and,
with the same statistic factors, reproduce the weights of the imaginary parts 
in eq.(13).
Only for the odd signature reggeons (the gluon) the counting is somewhat
different. First the contributions where both reggeons split into two gluons.
As we have discussed at the end of section 4, for an odd signature
reggeon the real part dominates, and according to (8) the diffractive cut with
zero multiplicity dominates. This leading contribution is given by the
BFKL Pomeron. In the nonleading QCD diagrams that we are presently studying we
expect to see the (subleading) imaginary parts of the gluons: with the 
symmetry properties
mentioned above it is easy to see that, in fact, they come in accordance with
(8). Finally the contributions where the reggeized gluon splits into three 
gluons. Let us consider the case where the cutting runs between line 1 and 2.
Obviously, the contribution 1(234) from above and below produces a diffractive
cut, the remaining three contributions lead to a cut reggeon. Similarly the
case where the cutting is between lines 3 and 4. Including sign and statistics
we get the weight -2/3! for the diffractive cut (for the explanation of the
negative sign see the argument before eq.(18)), and -2$\cdot$3/3! for the
contributions with a cut reggeon. From the term with cutting line 
between 2 and 3 we obtain
only cut reggeon graphs, with weight 4/$(2!)^2$. In the sum, all 
contributions with 
cut reggeons cancel, and we are left with the diffractive cut only: this is 
what we expect when we interpret these terms as three particle
contribution to the (odd signature) gluon trajectory function.\\ \\  
We are finally left with another type of contributions, the ``nondiagonal``
terms (Fig.16b). Here the reggeons from above and below do not match, and
these terms represent cut reggeon-reggeon interaction vertices for which the 
AGK arguments of the previous section do not apply. Some of these 
contributions are nothing but cuts (in external masses) of the BFKL kernel:
they are to be interpreted as another kind of ``bootstrap`` mechanism in
QCD. Other contributions represent new and higher order corrections to the
BFKL kernel: in  principle, it should be possible to derive in this way the 
corrections which have been studied in ~\cite{CW}. We hope to return this
question in a future paper, 
and we will not discuss it in any more detail here.
\\ \\
In summary, we have seen that all contributions obtained from the set of
QCD diagrams illustrated in Fig.14a fit into the AGK cutting rules; the 
only exception are the t-channel cuts of the four-reggeon vertices for which 
the AGK rules 
do not make any prediction. Compared to earlier studies, the new
features in QCD are the reggeization of the gluon and, more general, the
color degree of freedom. \\ \\
\section{Conclusion}
In this paper we have reviewed the ``old`` space-time picture of
multireggeon amplitudes and the physical interpretation of the AGK
cutting rules, almost forgotten during the last 10-15 years. Our main
interest was in the question of how these ideas are realized in
perturbative QCD. As to the space-time picture of the scattering
amplitude, our main new result is the appearance of a ``change of the
arrow direction``: an interaction between a pair of reggeons is
accompanied by a change of the direction of time (Fig.8b,d). One of the
reasons why, to our knowledge, this observation has not been made
before is that the t-channel iteration of two-ladder states has not
been studied very much; in QCD it was the calculation of the anomalous
dimension of the four-gluon operator which motivated a careful study
of these diagrams.\\ \\
Our study of the AGK cutting rules in QCD has, once more, emphasized the
importance of the reggeization of the gluon (bootstrap). In short,
the QCD diagrams which contribute to the present level of approximation
contain pieces which have to be counted as higher order corrections to
the BFKL ladder diagrams. They have to be removed before we are ready
to define four gluon states and the two-Pomeron exchange. As a result of this 
procedure, the (new) 2 $\to$ 4 gluon vertex emerges which has strikingly simple
symmetry properties. Also, there is no direct coupling of the four gluon state
to the quark pair, we are dealing with enhanced diagrams only.
After this decomposition into reggeizing and irreducible parts it is 
straightforward to verify the AGK cutting rules for the two-ladder 
exchanges. \\ \\
Moving into more detail of the AGK cutting rules, we have obtained a simple
matrix which describes, inside the Pomeron-Pomeron interaction,
the transition from one s-cut to another and which goes beyond the
original AGK rule in (13). Namely, in order to satisfy the AGK rules
the t-channel partial wave above and below the four gluon state
have to be the same, no matter whether the s-cut runs between the lines
1 and 2, 2 and 3, or 3 and 4. Once this partial wave by itself contains
two-reggeon states, it is a nontrivial question to ask how the weight
of the different cuts is preserved inside the partial
wave. Our answer to this question is contained in the matrix (19), and the
AGK weights appear as an eigenvector to this matrix. This is a new
result which has not been discussed before. \\ \\
In summary, we feel that the space-time analysis of these rather
complicated diagrams provides, despite al its subtleties, a fairly
easy picture of elastic scattering and diffractive dissociation
at high energies. It is also gratifying to see that the AGK rules are
satisfied, even if in QCD some details are quite unexpected.
As a next step, we believe, one should attempt to generalize the simple
probabilistic interpretation of the one-ladder approximation. For diagrams
with more than two t-channel lines one has, in addition to the familiar
squares of one-parton wave functions, also non-diagonal elements of a 
density matrix. The precise formulation of such a density matrix
requires many of the details presented in this paper.\\ \\
{\bf Acknowledgement:} One of us (M.R.) would like to thank DESY for
its kind hospitality and the Volkswagen-Stiftung for financial support.
\\ \\

\newpage
\begin{figure}
\begin{center}
\input paper27.fig1.pstex_t
\caption{(a) Single Ladder exchange; (b) Crossed ladder}
\input paper27.fig2.pstex_t
\caption{(a) Time ordering in the proton rest frame.The cutting line 
illustrates
an intermediate state in non-covariant perturbation theory; (b) time ordering 
in the cm-system}
\end{center}
\end{figure}
\vspace{1.0cm}
\begin{figure}
\begin{center}
\input paper27.fig3.pstex_t
\caption{Time ordering of the imaginary part of a single ladder:
(a) the energy discontinuity of Fig.1a; (b) the time ordering of the
energy discontinuity}
\input paper27.fig4.pstex_t
\caption{A reggeon diagram in QCD. The wavy lines denote reggeized gluons, the
black rung stands for the BFKL kernel. The right hand part illustrates the
space time structure of a reggeized gluon which, because of the bootstrap
property, can be redrawn as a ladder-like insertion}
\end{center}
\end{figure}
\begin{figure}
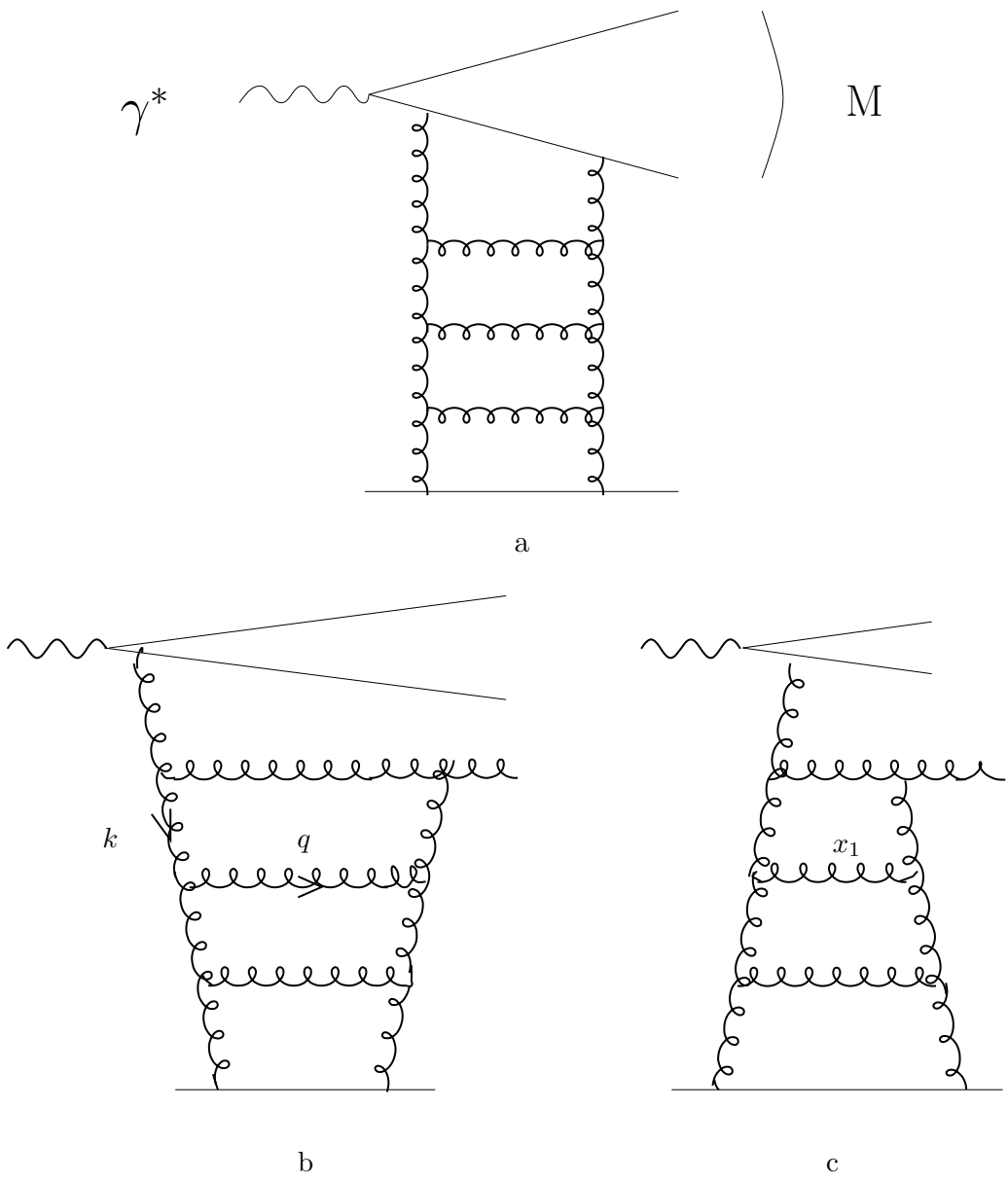

\begin{center}
\input paper27.fig5a.pstex_t
\input paper27.fig5b.pstex_t
\caption{(a) Single ladder exchange in diffractive dissociation (wavy lines 
denote gluons); (b)
time ordering in the target rest frame; (c) time ordering in the Breit 
frame}
\end{center}
\end{figure}
\begin{figure}
\begin{center}
\input paper27.fig6a.pstex_t
\vspace{0.5cm}
\input paper27.fig6b.pstex_t
\caption{Double ladder exchange (wavy lines denote gluons): (a) the simplest 
two-ladder diagram; (b) a more general example (enhanced diagram)}
\end{center}
\end{figure}
\begin{figure}
\begin{center}
\input paper27.fig7ab.pstex_t
\input paper27.fig7cd.pstex_t
\end{center}
\end{figure}
\begin{figure}
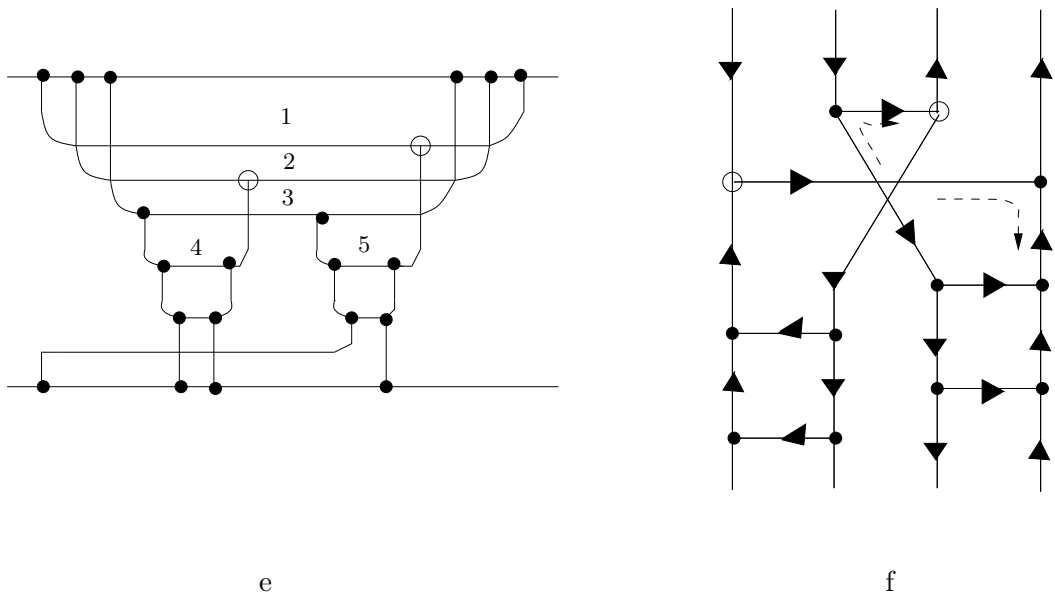

\begin{center}
\input paper27.fig7ef.pstex_t
\caption{Diagrams with four t-channel lines. (a) maximal nested structure at
the upper end; (b) example of a non-nested structure which does not contribute
to the leading log approximation; (c) example of a switch which vanishes at
high energies; (d) example of a switch for which a change of the time direction
on line 2 and 3 gives a nonvanishing contribution.
Vertices at which time changes its direction are marked by circles; 
(e) the space time structure of (d); (f) an example of a crossed diagram which contributes at high energies}
\end{center}
\end{figure}
\begin{figure}
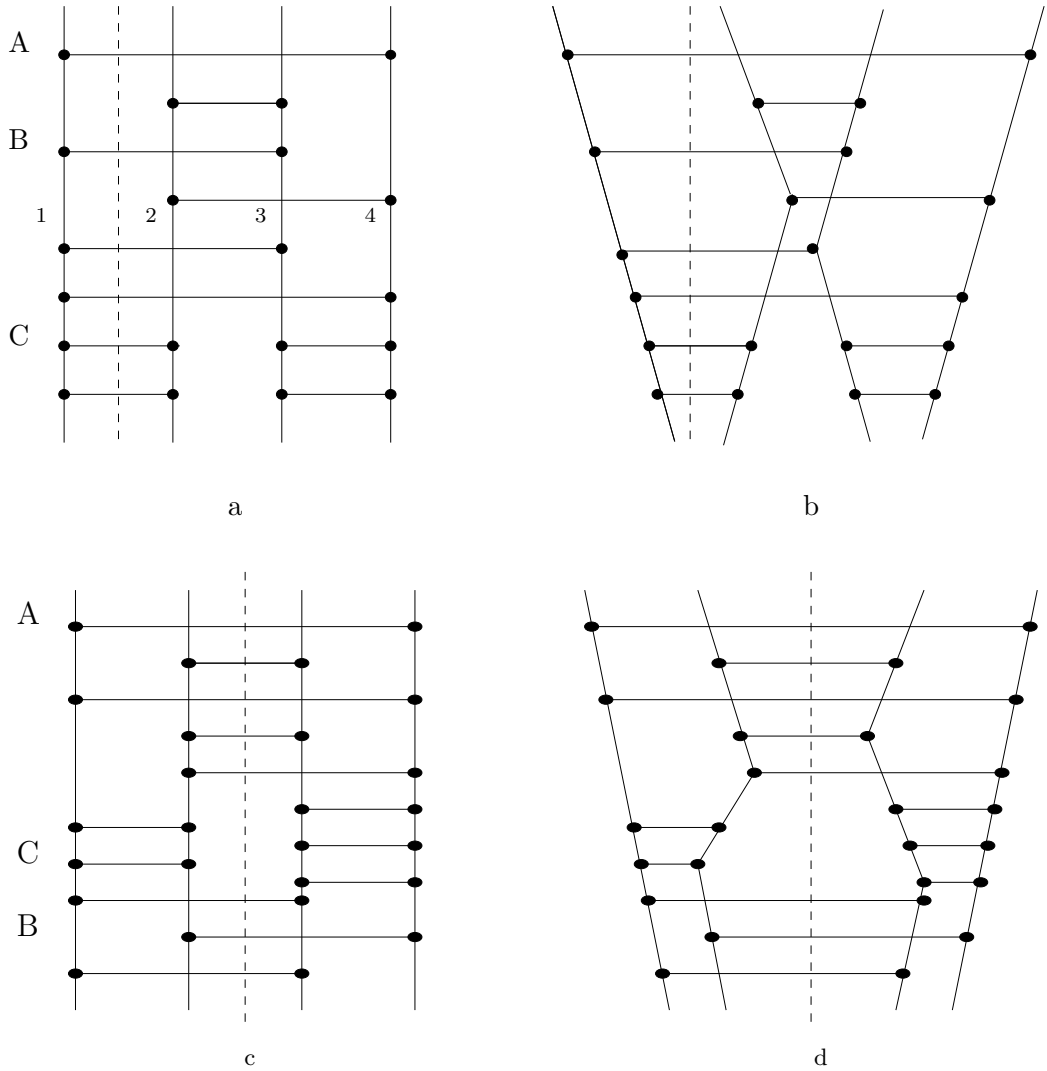

\begin{center}
\input paper27.fig8ab.pstex_t
\vspace{0.5cm}
\input paper27.fig8cd.pstex_t
\caption{Two more examples of nonvanishing contributions ((a) and (c)) with 
their space-time structure ((b) and (d), resp.}
\end{center}
\end{figure}
\begin{figure}
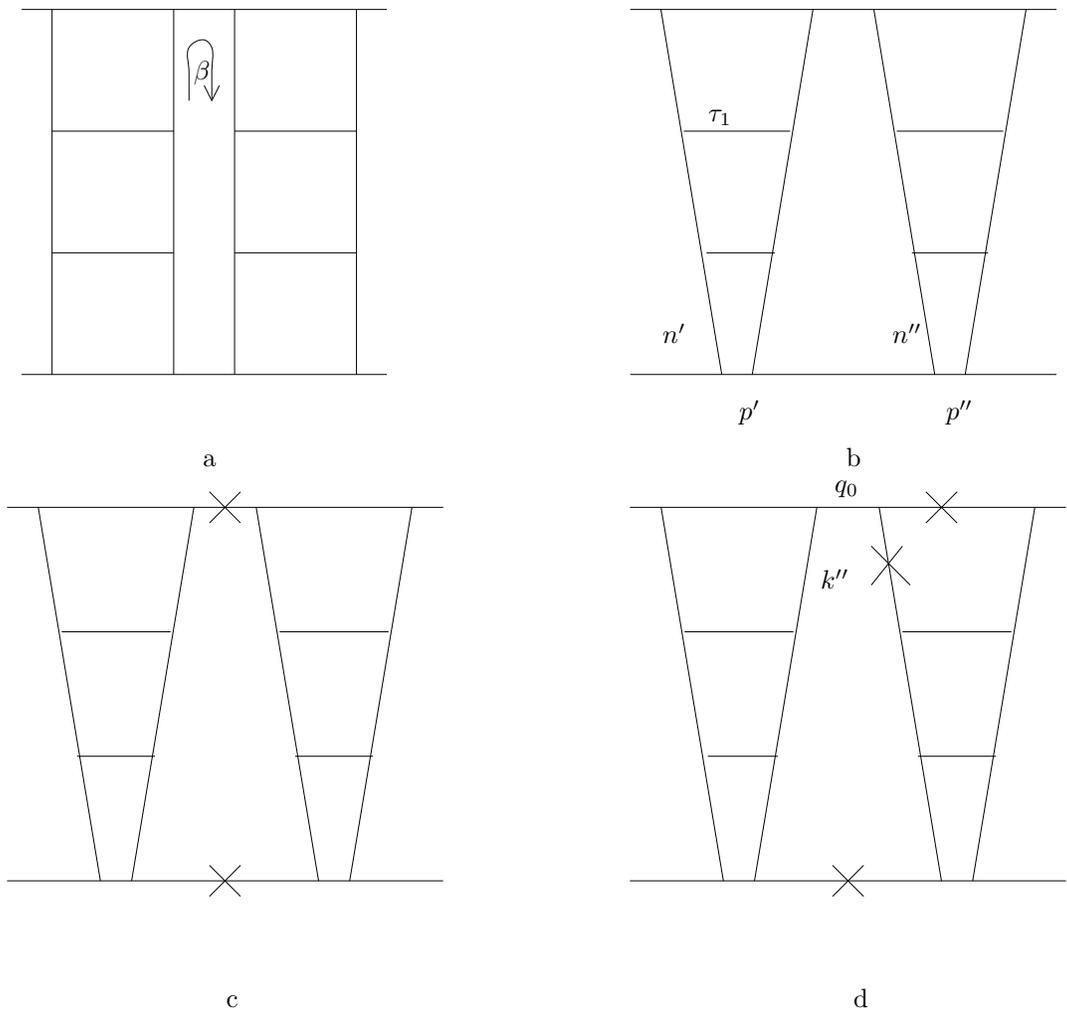

\begin{center}
\input paper27.fig9ab.pstex_t
\input paper27.fig9cd.pstex_t
\caption{The AFS diagram (planar coupling of two ladders to the upper line):
(a) the Feynman diagrams; (b) space time structure of (a);
(c) the elastic intermediate state; (d) another intermediate state, which
together with its hermitian conjugate, cancels against Fig.9c}
\end{center}
\end{figure}
\begin{figure}
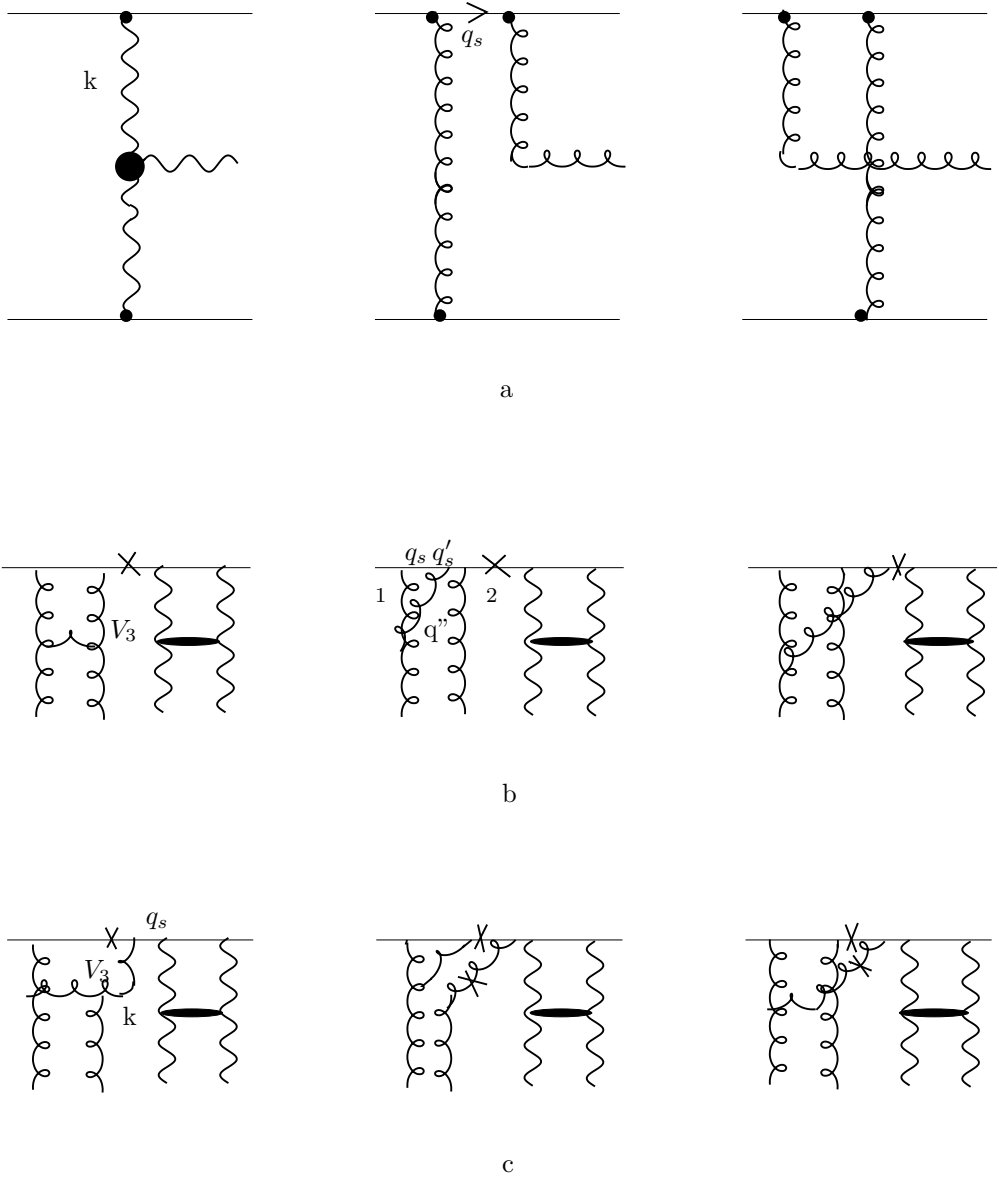

\begin{center}
\input paper27.fig10a.pstex_t
\input paper27.fig10b.pstex_t
\input paper27.fig10c.pstex_t
\caption{(a) Feynman diagrams which go into the BFKL production vertex;
(b) the single particle cut in the AFS diagram in QCD: wavy lines denote 
gluons, the black rung
the BFKL vertex. Inside the left hand side ladder, the BFKL rung has been 
resolved according to Fig.10a; (c) one of the two-particle cuts which cancels
against the single-particle cut in (b)}
\end{center}
\end{figure}
\begin{figure}
\begin{center}
\input paper27.fig11.pstex_t
\caption{Momentum structure of the triple gluon vertex}
\input paper27.fig12.pstex_t
\caption{A general two-reggeon exchange amplitude (wavy lines denote reggeons)}
\end{center}
\end{figure}
\begin{figure}
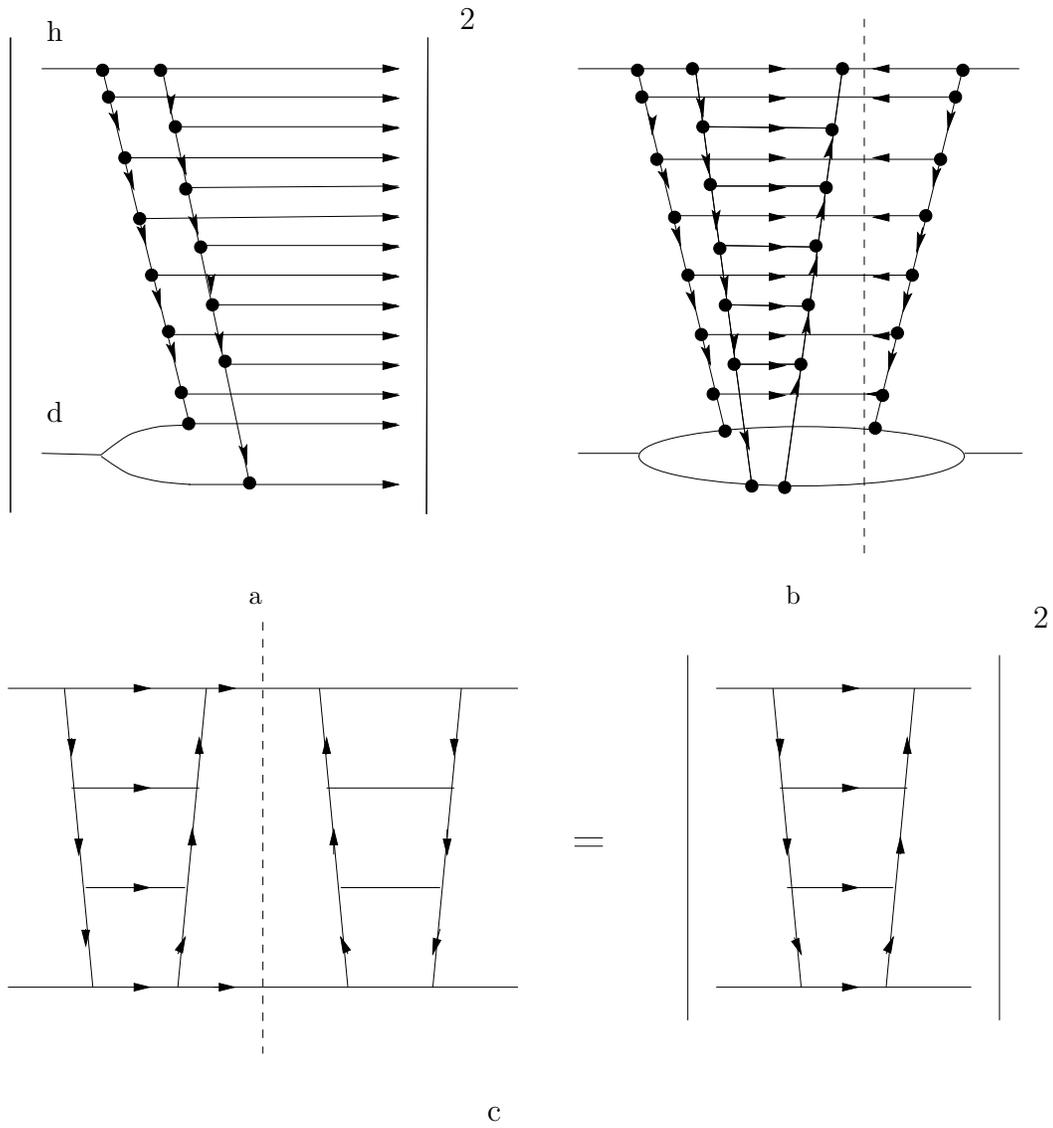

\begin{center}
\input paper27.fig13ab.pstex_t
\input paper27.fig13c.pstex_t
\caption{Space time structure of the corrections to the scattering of a hadron 
on a deuteron target: (a) final states with double multiplicity; (b) single
multiplicity final state; (c) the diffractive cut}
\end{center}
\end{figure}
\begin{figure}
\begin{center}
\input paper27.fig14a.pstex_t
\input paper27.fig14b.pstex_t
\end{center}
\end{figure}
\begin{figure}
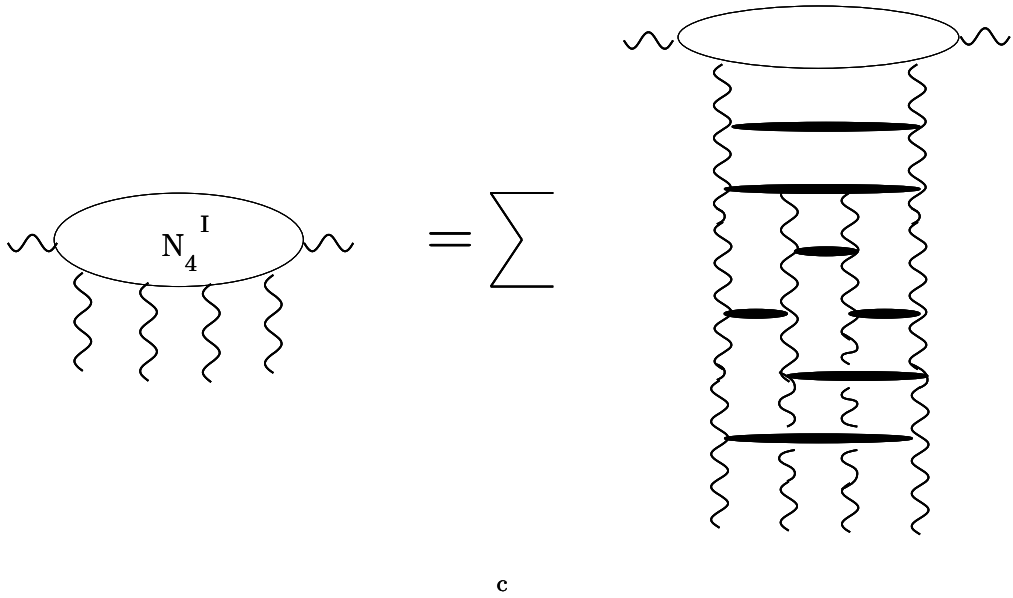

\begin{center}
\input paper27.fig14c.pstex_t
\caption{(a) General QCD gluon diagram with a four gluon state; after summing,
on the left hand side, over all contributions the sum of all diagrams can
be drawn in terms of effective (BFKL) vertices and reggeized gluons. The
discontinuity refers to angular momentum in the t-channel.
(b) decomposition of the partial wave $N_4$ with four reggeized gluons into 
the reggeizing part and the irreducible part; (c) illustration of the 
irreducible part}
\end{center}
\end{figure}
\begin{figure}
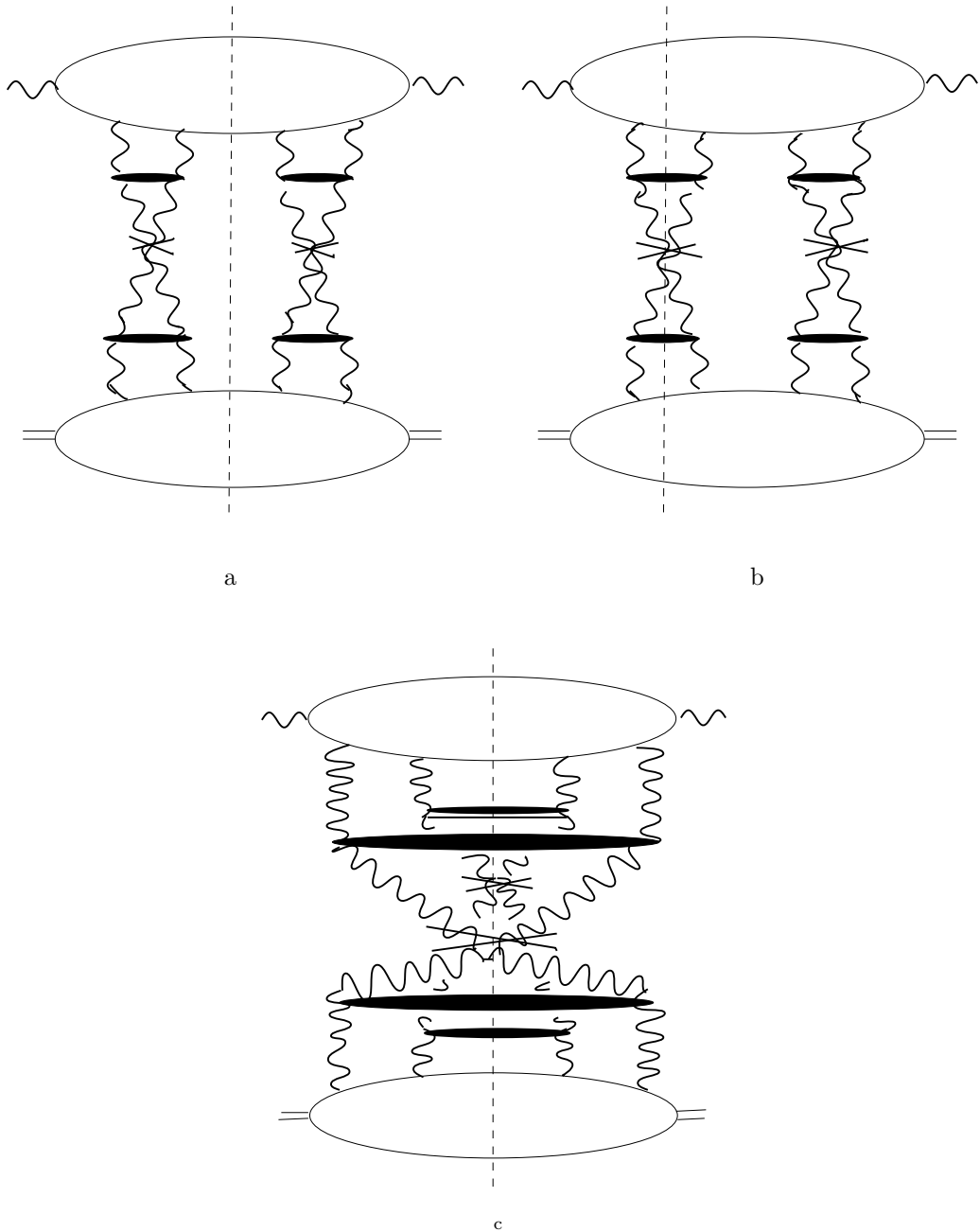

\begin{center}
\input paper27.fig15ab.pstex_t
\input paper27.fig15c.pstex_t
\caption{The AGK counting rules for the ladder states: (a) the diffractive 
cut $\sigma_0$; (b) the multiperipheral cut $\sigma_1$; (c) the double 
multiperipheral cut $\sigma_2$}
\end{center}
\end{figure}
\begin{figure}
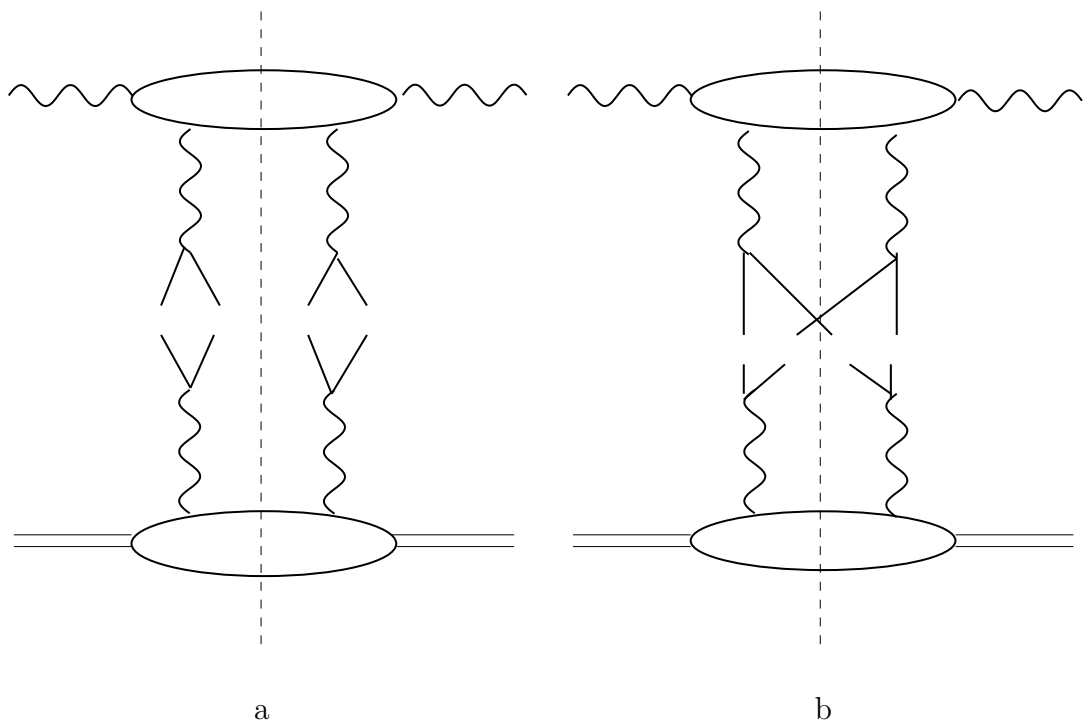

\begin{center}
\input paper27.fig16.pstex_t
\caption{Contribution of the reggizing pieces to the t-channel discontinuity.
(a) a ``diagonal term`` which matches eq.(8); (b) a ``nondiagonal term``
which is interpreted as a t-discontinuity of a four reggeon vertex}
\end{center}
\end{figure}
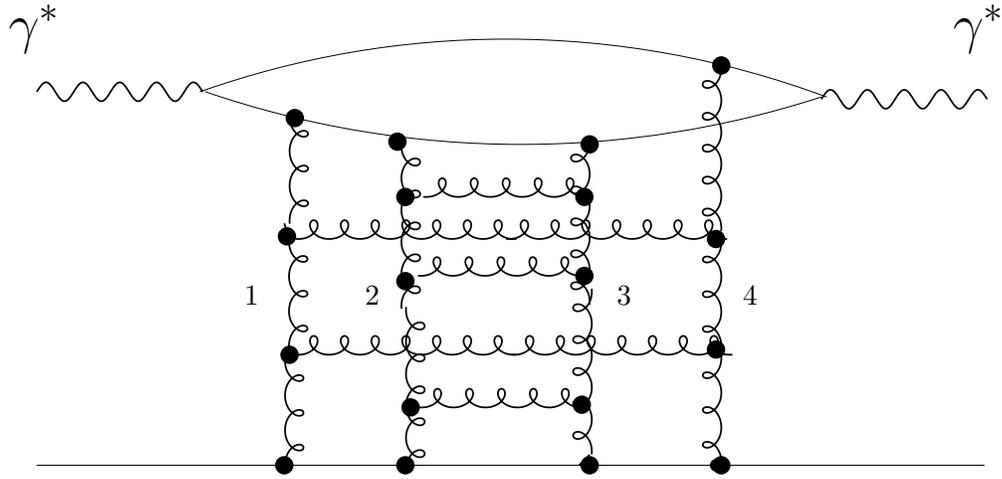
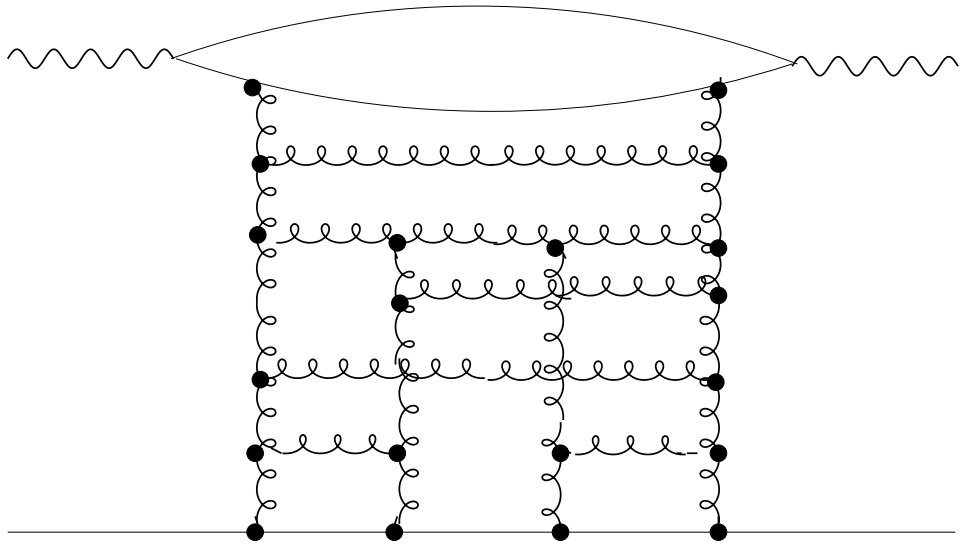
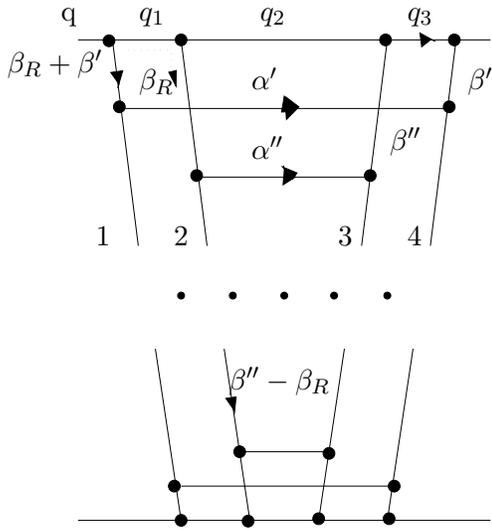
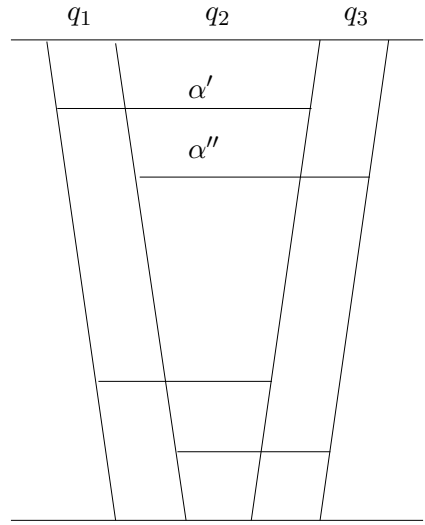
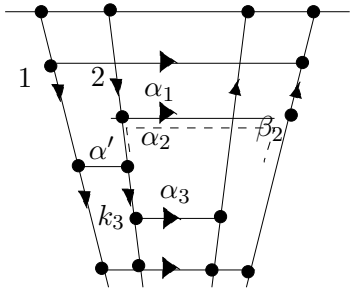
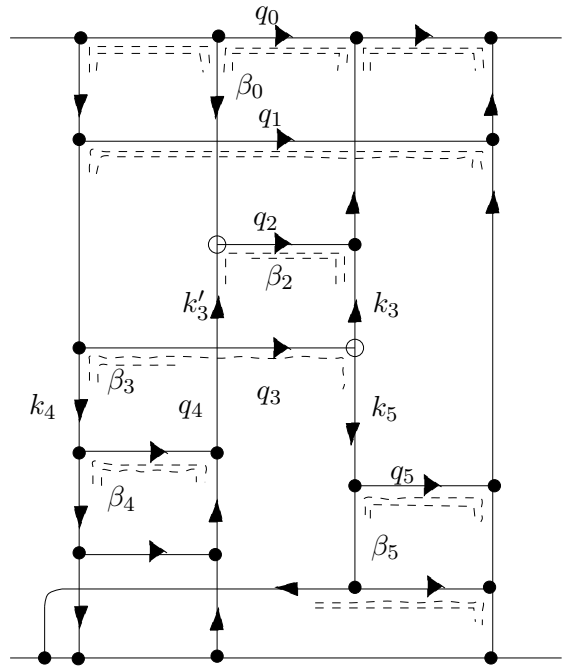
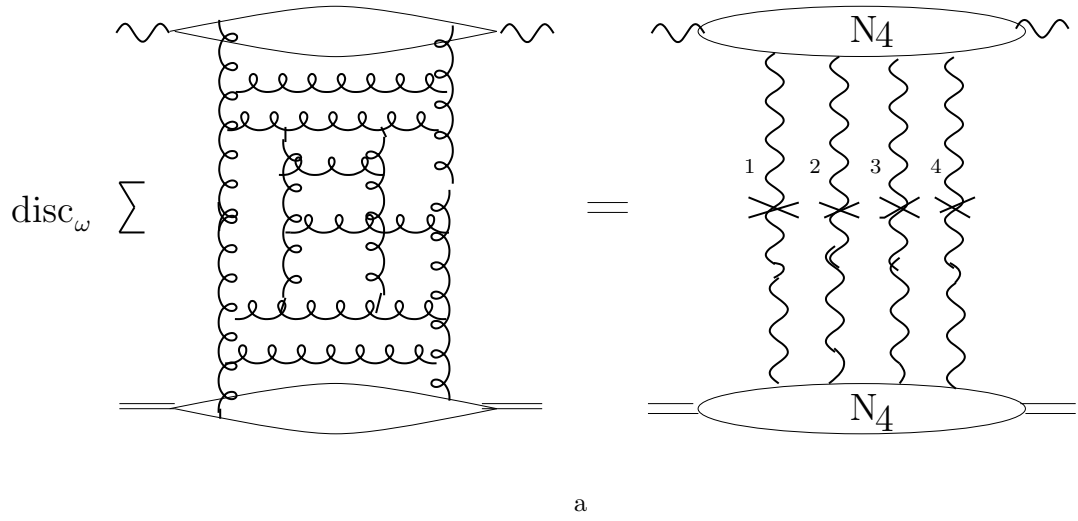
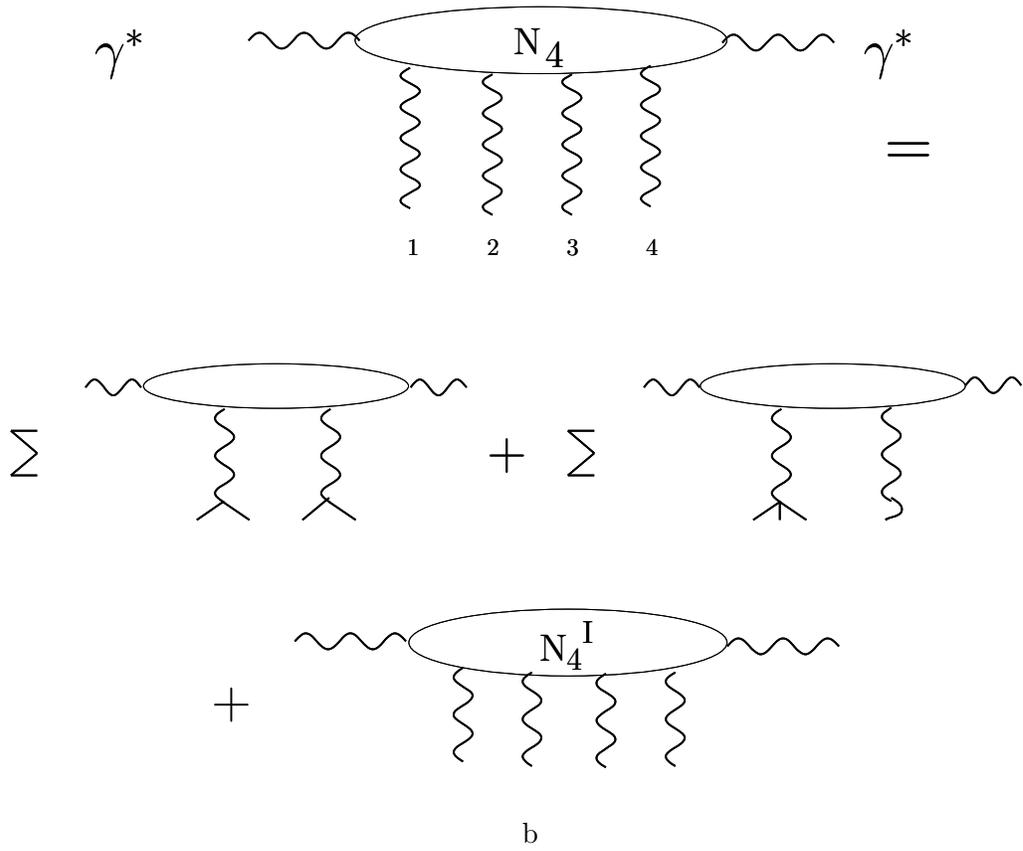
\end{document}

%% file: paper27.fig1.pstex_t
\begin{picture}(0,0)%
\epsfig{file=paper27.fig1.pstex}%
\end{picture}%
\setlength{\unitlength}{0.00066700in}%
\begingroup\makeatletter\ifx\SetFigFont\undefined
\def\x#1#2#3#4#5#6#7\relax{\def\x{#1#2#3#4#5#6}}%
\expandafter\x\fmtname xxxxxx\relax \def\y{splain}%
\ifx\x\y   
\gdef\SetFigFont#1#2#3{%
  \ifnum #1<17\tiny\else \ifnum #1<20\small\else
  \ifnum #1<24\normalsize\else \ifnum #1<29\large\else
  \ifnum #1<34\Large\else \ifnum #1<41\LARGE\else
     \huge\fi\fi\fi\fi\fi\fi
  \csname #3\endcsname}%
\else
\gdef\SetFigFont#1#2#3{\begingroup
  \count@#1\relax \ifnum 25<\count@\count@25\fi
  \def\x{\endgroup\@setsize\SetFigFont{#2pt}}%
  \expandafter\x
    \csname \romannumeral\the\count@ pt\expandafter\endcsname
    \csname @\romannumeral\the\count@ pt\endcsname
  \csname #3\endcsname}%
\fi
\fi\endgroup
\begin{picture}(7863,4236)(602,-4001)
\put(2261,-3973){\makebox(0,0)[lb]{\smash{\SetFigFont{10}{12.0}{rm}a}}}
\put(6856,-3973){\makebox(0,0)[lb]{\smash{\SetFigFont{10}{12.0}{rm}b}}}
\put(602,112){\makebox(0,0)[lb]{\smash{\SetFigFont{10}{12.0}{rm}q}}}
\put(729,-654){\makebox(0,0)[lb]{\smash{\SetFigFont{10}{12.0}{rm}$k_1$}}}
\put(2261,-1931){\makebox(0,0)[lb]{\smash{\SetFigFont{10}{12.0}{rm}$q_2$}}}
\put(2261,-2952){\makebox(0,0)[lb]{\smash{\SetFigFont{10}{12.0}{rm}$q_3$}}}
\put(3410,-654){\makebox(0,0)[lb]{\smash{\SetFigFont{10}{12.0}{rm}$k_1'$}}}
\put(602,-3080){\makebox(0,0)[lb]{\smash{\SetFigFont{10}{12.0}{rm}p}}}
\put(2261,-910){\makebox(0,0)[lb]{\smash{\SetFigFont{10}{12.0}{rm}$q_1$}}}
\put(729,-1675){\makebox(0,0)[lb]{\smash{\SetFigFont{10}{12.0}{rm}$k_2$}}}
\put(729,-2697){\makebox(0,0)[lb]{\smash{\SetFigFont{10}{12.0}{rm}$k_3$}}}
\put(3410,-1675){\makebox(0,0)[lb]{\smash{\SetFigFont{10}{12.0}{rm}$k_2'$}}}
\put(3410,-2697){\makebox(0,0)[lb]{\smash{\SetFigFont{10}{12.0}{rm}$k_3'$}}}
\end{picture}

%% file: paper27.fig2.pstex_t
\begin{picture}(0,0)%
\epsfig{file=paper27.fig2.pstex}%
\end{picture}%
\setlength{\unitlength}{0.00066700in}%
\begingroup\makeatletter\ifx\SetFigFont\undefined
\def\x#1#2#3#4#5#6#7\relax{\def\x{#1#2#3#4#5#6}}%
\expandafter\x\fmtname xxxxxx\relax \def\y{splain}%
\ifx\x\y   
\gdef\SetFigFont#1#2#3{%
  \ifnum #1<17\tiny\else \ifnum #1<20\small\else
  \ifnum #1<24\normalsize\else \ifnum #1<29\large\else
  \ifnum #1<34\Large\else \ifnum #1<41\LARGE\else
     \huge\fi\fi\fi\fi\fi\fi
  \csname #3\endcsname}%
\else
\gdef\SetFigFont#1#2#3{\begingroup
  \count@#1\relax \ifnum 25<\count@\count@25\fi
  \def\x{\endgroup\@setsize\SetFigFont{#2pt}}%
  \expandafter\x
    \csname \romannumeral\the\count@ pt\expandafter\endcsname
    \csname @\romannumeral\the\count@ pt\endcsname
  \csname #3\endcsname}%
\fi
\fi\endgroup
\begin{picture}(7801,4915)(603,-4457)
\put(603,-716){\makebox(0,0)[lb]{\smash{\SetFigFont{9}{10.8}{rm}t}}}
\put(4794,-716){\makebox(0,0)[lb]{\smash{\SetFigFont{9}{10.8}{rm}t}}}
\put(2382,-1786){\makebox(0,0)[lb]{\smash{\SetFigFont{10}{12.0}{rm}$q_2$}}}
\put(2443,-4425){\makebox(0,0)[lb]{\smash{\SetFigFont{12}{14.4}{rm}a}}}
\put(6617,-4425){\makebox(0,0)[lb]{\smash{\SetFigFont{12}{14.4}{rm}b}}}
\put(3151,314){\makebox(0,0)[lb]{\smash{\SetFigFont{10}{12.0}{rm}\vspace{0.5cm}}}}
\put(2382,-1295){\makebox(0,0)[lb]{\smash{\SetFigFont{10}{12.0}{rm}$q_1$}}}
\put(8404,-3976){\makebox(0,0)[lb]{\smash{\SetFigFont{9}{10.8}{rm}t}}}
\put(1154,-2032){\makebox(0,0)[lb]{\smash{\SetFigFont{10}{12.0}{rm}$t_2$}}}
\put(1031,-1418){\makebox(0,0)[lb]{\smash{\SetFigFont{10}{12.0}{rm}$t_1$}}}
\put(1154,-1725){\makebox(0,0)[lb]{\smash{\SetFigFont{10}{12.0}{rm}$k_2$}}}
\end{picture}

%% file: paper27.fig3.pstex_t
\begin{picture}(0,0)%
\epsfig{file=paper27.fig3.pstex}%
\end{picture}%
\setlength{\unitlength}{0.00066700in}%
\begingroup\makeatletter\ifx\SetFigFont\undefined
\def\x#1#2#3#4#5#6#7\relax{\def\x{#1#2#3#4#5#6}}%
\expandafter\x\fmtname xxxxxx\relax \def\y{splain}%
\ifx\x\y   
\gdef\SetFigFont#1#2#3{%
  \ifnum #1<17\tiny\else \ifnum #1<20\small\else
  \ifnum #1<24\normalsize\else \ifnum #1<29\large\else
  \ifnum #1<34\Large\else \ifnum #1<41\LARGE\else
     \huge\fi\fi\fi\fi\fi\fi
  \csname #3\endcsname}%
\else
\gdef\SetFigFont#1#2#3{\begingroup
  \count@#1\relax \ifnum 25<\count@\count@25\fi
  \def\x{\endgroup\@setsize\SetFigFont{#2pt}}%
  \expandafter\x
    \csname \romannumeral\the\count@ pt\expandafter\endcsname
    \csname @\romannumeral\the\count@ pt\endcsname
  \csname #3\endcsname}%
\fi
\fi\endgroup
\begin{picture}(5903,3136)(1266,-3646)
\put(2300,-2899){\makebox(0,0)[lb]{\smash{\SetFigFont{10}{12.0}{rm}x}}}
\put(2300,-2127){\makebox(0,0)[lb]{\smash{\SetFigFont{10}{12.0}{rm}x}}}
\put(2300,-1354){\makebox(0,0)[lb]{\smash{\SetFigFont{10}{12.0}{rm}x}}}
\put(2300,-583){\makebox(0,0)[lb]{\smash{\SetFigFont{10}{12.0}{rm}x}}}
\put(7076,-583){\makebox(0,0)[lb]{\smash{\SetFigFont{10}{12.0}{rm}x}}}
\put(3164,-1699){\makebox(0,0)[lb]{\smash{\SetFigFont{7}{8.4}{rm}$k_i'$}}}
\put(1324,-1699){\makebox(0,0)[lb]{\smash{\SetFigFont{7}{8.4}{rm}$k_i$}}}
\put(5956,-3621){\makebox(0,0)[lb]{\smash{\SetFigFont{10}{12.0}{rm}b}}}
\put(2253,-3621){\makebox(0,0)[lb]{\smash{\SetFigFont{10}{12.0}{rm}a}}}
\put(7076,-2899){\makebox(0,0)[lb]{\smash{\SetFigFont{10}{12.0}{rm}x}}}
\put(7076,-2127){\makebox(0,0)[lb]{\smash{\SetFigFont{10}{12.0}{rm}x}}}
\put(7076,-1354){\makebox(0,0)[lb]{\smash{\SetFigFont{10}{12.0}{rm}x}}}
\end{picture}

%% file: paper27.fig4.pstex_t
\begin{picture}(0,0)%
\epsfig{file=paper27.fig4.pstex}%
\end{picture}%
\setlength{\unitlength}{0.00066700in}%
\begingroup\makeatletter\ifx\SetFigFont\undefined
\def\x#1#2#3#4#5#6#7\relax{\def\x{#1#2#3#4#5#6}}%
\expandafter\x\fmtname xxxxxx\relax \def\y{splain}%
\ifx\x\y   
\gdef\SetFigFont#1#2#3{%
  \ifnum #1<17\tiny\else \ifnum #1<20\small\else
  \ifnum #1<24\normalsize\else \ifnum #1<29\large\else
  \ifnum #1<34\Large\else \ifnum #1<41\LARGE\else
     \huge\fi\fi\fi\fi\fi\fi
  \csname #3\endcsname}%
\else
\gdef\SetFigFont#1#2#3{\begingroup
  \count@#1\relax \ifnum 25<\count@\count@25\fi
  \def\x{\endgroup\@setsize\SetFigFont{#2pt}}%
  \expandafter\x
    \csname \romannumeral\the\count@ pt\expandafter\endcsname
    \csname @\romannumeral\the\count@ pt\endcsname
  \csname #3\endcsname}%
\fi
\fi\endgroup
\begin{picture}(6080,4335)(1191,-3877)
\put(4149,-2392){\makebox(0,0)[lb]{\smash{\SetFigFont{29}{34.8}{rm}=}}}
\put(5557,-3014){\makebox(0,0)[lb]{\smash{\SetFigFont{9}{10.8}{rm}$q_n$}}}
\put(3001,314){\makebox(0,0)[lb]{\smash{\SetFigFont{10}{12.0}{rm}\vspace{0.5cm}}}}
\end{picture}

%% file: paper27.fig5a.pstex_t
\begin{picture}(0,0)%
\epsfig{file=paper27.fig5a.pstex}%
\end{picture}%
\setlength{\unitlength}{0.00070000in}%
\begingroup\makeatletter\ifx\SetFigFont\undefined
\def\x#1#2#3#4#5#6#7\relax{\def\x{#1#2#3#4#5#6}}%
\expandafter\x\fmtname xxxxxx\relax \def\y{splain}%
\ifx\x\y   
\gdef\SetFigFont#1#2#3{%
  \ifnum #1<17\tiny\else \ifnum #1<20\small\else
  \ifnum #1<24\normalsize\else \ifnum #1<29\large\else
  \ifnum #1<34\Large\else \ifnum #1<41\LARGE\else
     \huge\fi\fi\fi\fi\fi\fi
  \csname #3\endcsname}%
\else
\gdef\SetFigFont#1#2#3{\begingroup
  \count@#1\relax \ifnum 25<\count@\count@25\fi
  \def\x{\endgroup\@setsize\SetFigFont{#2pt}}%
  \expandafter\x
    \csname \romannumeral\the\count@ pt\expandafter\endcsname
    \csname @\romannumeral\the\count@ pt\endcsname
  \csname #3\endcsname}%
\fi
\fi\endgroup
\begin{picture}(5915,4472)(1803,-4701)
\put(4824,-4678){\makebox(0,0)[lb]{\smash{\SetFigFont{7}{8.4}{rm}\vspace{0.5cm}}}}
\put(4824,-4389){\makebox(0,0)[lb]{\smash{\SetFigFont{12}{14.4}{rm}a}}}
\put(1803,-1109){\makebox(0,0)[lb]{\smash{\SetFigFont{20}{24.0}{rm}$\gamma^*$}}}
\put(7364,-1045){\makebox(0,0)[lb]{\smash{\SetFigFont{17}{20.4}{rm}M}}}
\end{picture}

%% file: paper27.fig5b.pstex_t
\begin{picture}(0,0)%
\epsfig{file=paper27.fig5b.pstex}%
\end{picture}%
\setlength{\unitlength}{0.00070000in}%
\begingroup\makeatletter\ifx\SetFigFont\undefined
\def\x#1#2#3#4#5#6#7\relax{\def\x{#1#2#3#4#5#6}}%
\expandafter\x\fmtname xxxxxx\relax \def\y{splain}%
\ifx\x\y   
\gdef\SetFigFont#1#2#3{%
  \ifnum #1<17\tiny\else \ifnum #1<20\small\else
  \ifnum #1<24\normalsize\else \ifnum #1<29\large\else
  \ifnum #1<34\Large\else \ifnum #1<41\LARGE\else
     \huge\fi\fi\fi\fi\fi\fi
  \csname #3\endcsname}%
\else
\gdef\SetFigFont#1#2#3{\begingroup
  \count@#1\relax \ifnum 25<\count@\count@25\fi
  \def\x{\endgroup\@setsize\SetFigFont{#2pt}}%
  \expandafter\x
    \csname \romannumeral\the\count@ pt\expandafter\endcsname
    \csname @\romannumeral\the\count@ pt\endcsname
  \csname #3\endcsname}%
\fi
\fi\endgroup
\begin{picture}(7700,4451)(491,-4996)
\put(2740,-2476){\makebox(0,0)[lb]{\smash{\SetFigFont{11}{13.2}{rm}$q$}}}
\put(6847,-2517){\makebox(0,0)[lb]{\smash{\SetFigFont{11}{13.2}{rm}$x_1$}}}
\put(2746,-4966){\makebox(0,0)[lb]{\smash{\SetFigFont{11}{13.2}{rm}b}}}
\put(6796,-4966){\makebox(0,0)[lb]{\smash{\SetFigFont{11}{13.2}{rm}c}}}
\put(1250,-2476){\makebox(0,0)[lb]{\smash{\SetFigFont{11}{13.2}{rm}$k$}}}
\end{picture}

%% file: paper27.fig6a.pstex_t
\begin{picture}(0,0)%
\epsfig{file=paper27.fig6a.pstex}%
\end{picture}%
\setlength{\unitlength}{0.00061200in}%
\begingroup\makeatletter\ifx\SetFigFont\undefined
\def\x#1#2#3#4#5#6#7\relax{\def\x{#1#2#3#4#5#6}}%
\expandafter\x\fmtname xxxxxx\relax \def\y{splain}%
\ifx\x\y   
\gdef\SetFigFont#1#2#3{%
  \ifnum #1<17\tiny\else \ifnum #1<20\small\else
  \ifnum #1<24\normalsize\else \ifnum #1<29\large\else
  \ifnum #1<34\Large\else \ifnum #1<41\LARGE\else
     \huge\fi\fi\fi\fi\fi\fi
  \csname #3\endcsname}%
\else
\gdef\SetFigFont#1#2#3{\begingroup
  \count@#1\relax \ifnum 25<\count@\count@25\fi
  \def\x{\endgroup\@setsize\SetFigFont{#2pt}}%
  \expandafter\x
    \csname \romannumeral\the\count@ pt\expandafter\endcsname
    \csname @\romannumeral\the\count@ pt\endcsname
  \csname #3\endcsname}%
\fi
\fi\endgroup
\begin{picture}(8363,5055)(676,-4957)
\put(5896,-2536){\makebox(0,0)[lb]{\smash{\SetFigFont{11}{13.2}{rm}3}}}
\put(6976,-2536){\makebox(0,0)[lb]{\smash{\SetFigFont{11}{13.2}{rm}4}}}
\put(2701,-2536){\makebox(0,0)[lb]{\smash{\SetFigFont{11}{13.2}{rm}1}}}
\put(3736,-2536){\makebox(0,0)[lb]{\smash{\SetFigFont{11}{13.2}{rm}2}}}
\put(4906,-4921){\makebox(0,0)[lb]{\smash{\SetFigFont{9}{10.8}{rm}\vspace{0.5cm}}}}
\put(8776,-286){\makebox(0,0)[lb]{\smash{\SetFigFont{20}{24.0}{rm}$\gamma^*$}}}
\put(4906,-4561){\makebox(0,0)[lb]{\smash{\SetFigFont{14}{16.8}{rm}a}}}
\put(676,-286){\makebox(0,0)[lb]{\smash{\SetFigFont{20}{24.0}{rm}$\gamma^*$}}}
\end{picture}

%% file: paper27.fig6b.pstex_t
\begin{picture}(0,0)%
\epsfig{file=paper27.fig6b.pstex}%
\end{picture}%
\setlength{\unitlength}{0.00061200in}%
\begingroup\makeatletter\ifx\SetFigFont\undefined
\def\x#1#2#3#4#5#6#7\relax{\def\x{#1#2#3#4#5#6}}%
\expandafter\x\fmtname xxxxxx\relax \def\y{splain}%
\ifx\x\y   
\gdef\SetFigFont#1#2#3{%
  \ifnum #1<17\tiny\else \ifnum #1<20\small\else
  \ifnum #1<24\normalsize\else \ifnum #1<29\large\else
  \ifnum #1<34\Large\else \ifnum #1<41\LARGE\else
     \huge\fi\fi\fi\fi\fi\fi
  \csname #3\endcsname}%
\else
\gdef\SetFigFont#1#2#3{\begingroup
  \count@#1\relax \ifnum 25<\count@\count@25\fi
  \def\x{\endgroup\@setsize\SetFigFont{#2pt}}%
  \expandafter\x
    \csname \romannumeral\the\count@ pt\expandafter\endcsname
    \csname @\romannumeral\the\count@ pt\endcsname
  \csname #3\endcsname}%
\fi
\fi\endgroup
\begin{picture}(8153,5218)(886,-5494)
\put(4951,-5461){\makebox(0,0)[lb]{\smash{\SetFigFont{11}{13.2}{rm}b}}}
\end{picture}

%% file: paper27.fig7ab.pstex_t
\begin{picture}(0,0)%
\epsfig{file=paper27.fig7ab.pstex}%
\end{picture}%
\setlength{\unitlength}{0.00066700in}%
\begingroup\makeatletter\ifx\SetFigFont\undefined
\def\x#1#2#3#4#5#6#7\relax{\def\x{#1#2#3#4#5#6}}%
\expandafter\x\fmtname xxxxxx\relax \def\y{splain}%
\ifx\x\y   
\gdef\SetFigFont#1#2#3{%
  \ifnum #1<17\tiny\else \ifnum #1<20\small\else
  \ifnum #1<24\normalsize\else \ifnum #1<29\large\else
  \ifnum #1<34\Large\else \ifnum #1<41\LARGE\else
     \huge\fi\fi\fi\fi\fi\fi
  \csname #3\endcsname}%
\else
\gdef\SetFigFont#1#2#3{\begingroup
  \count@#1\relax \ifnum 25<\count@\count@25\fi
  \def\x{\endgroup\@setsize\SetFigFont{#2pt}}%
  \expandafter\x
    \csname \romannumeral\the\count@ pt\expandafter\endcsname
    \csname @\romannumeral\the\count@ pt\endcsname
  \csname #3\endcsname}%
\fi
\fi\endgroup
\begin{picture}(8120,4647)(751,-4735)
\put(2723,-256){\makebox(0,0)[lb]{\smash{\SetFigFont{11}{13.2}{rm}$q_2$}}}
\put(1779,-256){\makebox(0,0)[lb]{\smash{\SetFigFont{11}{13.2}{rm}$q_1$}}}
\put(1172,-256){\makebox(0,0)[lb]{\smash{\SetFigFont{11}{13.2}{rm}q}}}
\put(3870,-256){\makebox(0,0)[lb]{\smash{\SetFigFont{11}{13.2}{rm}$q_3$}}}
\put(3734,-1267){\makebox(0,0)[lb]{\smash{\SetFigFont{11}{13.2}{rm}$\beta''$}}}
\put(1441,-2009){\makebox(0,0)[lb]{\smash{\SetFigFont{11}{13.2}{rm}1}}}
\put(2049,-2009){\makebox(0,0)[lb]{\smash{\SetFigFont{11}{13.2}{rm}2}}}
\put(3330,-2009){\makebox(0,0)[lb]{\smash{\SetFigFont{11}{13.2}{rm}3}}}
\put(2476,-3136){\makebox(0,0)[lb]{\smash{\SetFigFont{11}{13.2}{rm}$\beta''-\beta_R$}}}
\put(751,-661){\makebox(0,0)[lb]{\smash{\SetFigFont{11}{13.2}{rm}$\beta_R + \beta'$}}}
\put(1779,-795){\makebox(0,0)[lb]{\smash{\SetFigFont{11}{13.2}{rm}$\beta_R$}}}
\put(3870,-2009){\makebox(0,0)[lb]{\smash{\SetFigFont{11}{13.2}{rm}4}}}
\put(4341,-795){\makebox(0,0)[lb]{\smash{\SetFigFont{11}{13.2}{rm}$\beta'$}}}
\put(7106,-256){\makebox(0,0)[lb]{\smash{\SetFigFont{11}{13.2}{rm}$q_2$}}}
\put(6027,-256){\makebox(0,0)[lb]{\smash{\SetFigFont{11}{13.2}{rm}$q_1$}}}
\put(7241,-4705){\makebox(0,0)[lb]{\smash{\SetFigFont{11}{13.2}{rm}b}}}
\put(2925,-4705){\makebox(0,0)[lb]{\smash{\SetFigFont{11}{13.2}{rm}a}}}
\put(8185,-256){\makebox(0,0)[lb]{\smash{\SetFigFont{11}{13.2}{rm}$q_3$}}}
\put(2655,-1335){\makebox(0,0)[lb]{\smash{\SetFigFont{11}{13.2}{rm}$\alpha''$}}}
\put(2655,-795){\makebox(0,0)[lb]{\smash{\SetFigFont{11}{13.2}{rm}$\alpha'$}}}
\put(6972,-1335){\makebox(0,0)[lb]{\smash{\SetFigFont{11}{13.2}{rm}$\alpha''$}}}
\put(6972,-863){\makebox(0,0)[lb]{\smash{\SetFigFont{11}{13.2}{rm}$\alpha'$}}}
\end{picture}

%% file: paper27.fig7cd.pstex_t
\begin{picture}(0,0)%
\epsfig{file=paper27.fig7cd.pstex}%
\end{picture}%
\setlength{\unitlength}{0.00066700in}%
\begingroup\makeatletter\ifx\SetFigFont\undefined
\def\x#1#2#3#4#5#6#7\relax{\def\x{#1#2#3#4#5#6}}%
\expandafter\x\fmtname xxxxxx\relax \def\y{splain}%
\ifx\x\y   
\gdef\SetFigFont#1#2#3{%
  \ifnum #1<17\tiny\else \ifnum #1<20\small\else
  \ifnum #1<24\normalsize\else \ifnum #1<29\large\else
  \ifnum #1<34\Large\else \ifnum #1<41\LARGE\else
     \huge\fi\fi\fi\fi\fi\fi
  \csname #3\endcsname}%
\else
\gdef\SetFigFont#1#2#3{\begingroup
  \count@#1\relax \ifnum 25<\count@\count@25\fi
  \def\x{\endgroup\@setsize\SetFigFont{#2pt}}%
  \expandafter\x
    \csname \romannumeral\the\count@ pt\expandafter\endcsname
    \csname @\romannumeral\the\count@ pt\endcsname
  \csname #3\endcsname}%
\fi
\fi\endgroup
\begin{picture}(7871,6746)(574,-6622)
\put(6917,-2330){\makebox(0,0)[lb]{\smash{\SetFigFont{11}{13.2}{rm}$k_3$}}}
\put(5402,-3116){\makebox(0,0)[lb]{\smash{\SetFigFont{11}{13.2}{rm}$q_4$}}}
\put(4239,-3128){\makebox(0,0)[lb]{\smash{\SetFigFont{11}{13.2}{rm}$k_4$}}}
\put(6012,-3007){\makebox(0,0)[lb]{\smash{\SetFigFont{11}{13.2}{rm}$q_3$}}}
\put(1249,-2411){\makebox(0,0)[lb]{\smash{\SetFigFont{11}{13.2}{rm}$\alpha'$}}}
\put(5998,-44){\makebox(0,0)[lb]{\smash{\SetFigFont{11}{13.2}{rm}$q_0$}}}
\put(6025,-842){\makebox(0,0)[lb]{\smash{\SetFigFont{11}{13.2}{rm}$q_1$}}}
\put(5984,-1653){\makebox(0,0)[lb]{\smash{\SetFigFont{11}{13.2}{rm}$q_2$}}}
\put(5430,-2330){\makebox(0,0)[lb]{\smash{\SetFigFont{11}{13.2}{rm}$k_3'$}}}
\put(2551,-2236){\makebox(0,0)[lb]{\smash{\SetFigFont{11}{13.2}{rm}$\beta_2$}}}
\put(4847,-3846){\makebox(0,0)[lb]{\smash{\SetFigFont{11}{13.2}{rm}$\beta_4$}}}
\put(4847,-2926){\makebox(0,0)[lb]{\smash{\SetFigFont{11}{13.2}{rm}$\beta_3$}}}
\put(6905,-4225){\makebox(0,0)[lb]{\smash{\SetFigFont{11}{13.2}{rm}$\beta_5$}}}
\put(6905,-3142){\makebox(0,0)[lb]{\smash{\SetFigFont{11}{13.2}{rm}$k_5$}}}
\put(7053,-3630){\makebox(0,0)[lb]{\smash{\SetFigFont{11}{13.2}{rm}$q_5$}}}
\put(5849,-625){\makebox(0,0)[lb]{\smash{\SetFigFont{11}{13.2}{rm}$\beta_0$}}}
\put(6079,-2114){\makebox(0,0)[lb]{\smash{\SetFigFont{11}{13.2}{rm}$\beta_2$}}}
\put(1316,-2926){\makebox(0,0)[lb]{\smash{\SetFigFont{11}{13.2}{rm}$k_3$}}}
\put(1804,-2723){\makebox(0,0)[lb]{\smash{\SetFigFont{11}{13.2}{rm}$\alpha_3$}}}
\put(1655,-2304){\makebox(0,0)[lb]{\smash{\SetFigFont{11}{13.2}{rm}$\alpha_2$}}}
\put(1682,-1925){\makebox(0,0)[lb]{\smash{\SetFigFont{11}{13.2}{rm}$\alpha_1$}}}
\put(1263,-1843){\makebox(0,0)[lb]{\smash{\SetFigFont{11}{13.2}{rm}2}}}
\put(4426,-6586){\makebox(0,0)[lb]{\smash{\SetFigFont{10}{12.0}{rm}\vspace{0.5cm}}}}
\put(1884,-5524){\makebox(0,0)[lb]{\smash{\SetFigFont{11}{13.2}{rm}c}}}
\put(6242,-5551){\makebox(0,0)[lb]{\smash{\SetFigFont{11}{13.2}{rm}d}}}
\put(695,-1857){\makebox(0,0)[lb]{\smash{\SetFigFont{11}{13.2}{rm}1}}}
\end{picture}

%% file: paper27.fig7ef.pstex_t
\begin{picture}(0,0)%
\epsfig{file=paper27.fig7ef.pstex}%
\end{picture}%
\setlength{\unitlength}{0.00066700in}%
\begingroup\makeatletter\ifx\SetFigFont\undefined
\def\x#1#2#3#4#5#6#7\relax{\def\x{#1#2#3#4#5#6}}%
\expandafter\x\fmtname xxxxxx\relax \def\y{splain}%
\ifx\x\y   
\gdef\SetFigFont#1#2#3{%
  \ifnum #1<17\tiny\else \ifnum #1<20\small\else
  \ifnum #1<24\normalsize\else \ifnum #1<29\large\else
  \ifnum #1<34\Large\else \ifnum #1<41\LARGE\else
     \huge\fi\fi\fi\fi\fi\fi
  \csname #3\endcsname}%
\else
\gdef\SetFigFont#1#2#3{\begingroup
  \count@#1\relax \ifnum 25<\count@\count@25\fi
  \def\x{\endgroup\@setsize\SetFigFont{#2pt}}%
  \expandafter\x
    \csname \romannumeral\the\count@ pt\expandafter\endcsname
    \csname @\romannumeral\the\count@ pt\endcsname
  \csname #3\endcsname}%
\fi
\fi\endgroup
\begin{picture}(8227,4624)(740,-4577)
\put(7609,-4547){\makebox(0,0)[lb]{\smash{\SetFigFont{11}{13.2}{rm}f}}}
\put(2726,-4547){\makebox(0,0)[lb]{\smash{\SetFigFont{11}{13.2}{rm}e}}}
\put(3498,-1897){\makebox(0,0)[lb]{\smash{\SetFigFont{9}{10.8}{rm}5}}}
\put(2186,-1925){\makebox(0,0)[lb]{\smash{\SetFigFont{9}{10.8}{rm}4}}}
\put(2903,-1533){\makebox(0,0)[lb]{\smash{\SetFigFont{9}{10.8}{rm}3}}}
\put(2915,-1249){\makebox(0,0)[lb]{\smash{\SetFigFont{9}{10.8}{rm}2}}}
\put(2889,-898){\makebox(0,0)[lb]{\smash{\SetFigFont{9}{10.8}{rm}1}}}
\end{picture}

%% file: paper27.fig8ab.pstex_t
\begin{picture}(0,0)%
\epsfig{file=paper27.fig8ab.pstex}%
\end{picture}%
\setlength{\unitlength}{0.00066700in}%
\begingroup\makeatletter\ifx\SetFigFont\undefined
\def\x#1#2#3#4#5#6#7\relax{\def\x{#1#2#3#4#5#6}}%
\expandafter\x\fmtname xxxxxx\relax \def\y{splain}%
\ifx\x\y   
\gdef\SetFigFont#1#2#3{%
  \ifnum #1<17\tiny\else \ifnum #1<20\small\else
  \ifnum #1<24\normalsize\else \ifnum #1<29\large\else
  \ifnum #1<34\Large\else \ifnum #1<41\LARGE\else
     \huge\fi\fi\fi\fi\fi\fi
  \csname #3\endcsname}%
\else
\gdef\SetFigFont#1#2#3{\begingroup
  \count@#1\relax \ifnum 25<\count@\count@25\fi
  \def\x{\endgroup\@setsize\SetFigFont{#2pt}}%
  \expandafter\x
    \csname \romannumeral\the\count@ pt\expandafter\endcsname
    \csname @\romannumeral\the\count@ pt\endcsname
  \csname #3\endcsname}%
\fi
\fi\endgroup
\begin{picture}(8134,4397)(677,-4972)
\put(3455,-2302){\makebox(0,0)[lb]{\smash{\SetFigFont{8}{9.6}{rm}4}}}
\put(2601,-2302){\makebox(0,0)[lb]{\smash{\SetFigFont{8}{9.6}{rm}3}}}
\put(1746,-2302){\makebox(0,0)[lb]{\smash{\SetFigFont{8}{9.6}{rm}2}}}
\put(889,-2302){\makebox(0,0)[lb]{\smash{\SetFigFont{8}{9.6}{rm}1}}}
\put(2387,-4590){\makebox(0,0)[lb]{\smash{\SetFigFont{11}{13.2}{rm}a}}}
\put(3976,-4936){\makebox(0,0)[lb]{\smash{\SetFigFont{10}{12.0}{rm}\vspace{0.5cm}}}}
\put(677,-3256){\makebox(0,0)[lb]{\smash{\SetFigFont{11}{13.2}{rm}C}}}
\put(677,-1731){\makebox(0,0)[lb]{\smash{\SetFigFont{11}{13.2}{rm}B}}}
\put(677,-969){\makebox(0,0)[lb]{\smash{\SetFigFont{11}{13.2}{rm}A}}}
\put(6875,-4590){\makebox(0,0)[lb]{\smash{\SetFigFont{11}{13.2}{rm}b}}}
\end{picture}

%% file: paper27.fig8cd.pstex_t
\begin{picture}(0,0)%
\epsfig{file=paper27.fig8cd.pstex}%
\end{picture}%
\setlength{\unitlength}{0.00066700in}%
\begingroup\makeatletter\ifx\SetFigFont\undefined
\def\x#1#2#3#4#5#6#7\relax{\def\x{#1#2#3#4#5#6}}%
\expandafter\x\fmtname xxxxxx\relax \def\y{splain}%
\ifx\x\y   
\gdef\SetFigFont#1#2#3{%
  \ifnum #1<17\tiny\else \ifnum #1<20\small\else
  \ifnum #1<24\normalsize\else \ifnum #1<29\large\else
  \ifnum #1<34\Large\else \ifnum #1<41\LARGE\else
     \huge\fi\fi\fi\fi\fi\fi
  \csname #3\endcsname}%
\else
\gdef\SetFigFont#1#2#3{\begingroup
  \count@#1\relax \ifnum 25<\count@\count@25\fi
  \def\x{\endgroup\@setsize\SetFigFont{#2pt}}%
  \expandafter\x
    \csname \romannumeral\the\count@ pt\expandafter\endcsname
    \csname @\romannumeral\the\count@ pt\endcsname
  \csname #3\endcsname}%
\fi
\fi\endgroup
\begin{picture}(8007,3907)(602,-4472)
\put(602,-2873){\makebox(0,0)[lb]{\smash{\SetFigFont{12}{14.4}{rm}C}}}
\put(602,-3446){\makebox(0,0)[lb]{\smash{\SetFigFont{12}{14.4}{rm}B}}}
\put(6821,-4451){\makebox(0,0)[lb]{\smash{\SetFigFont{9}{10.8}{rm}d}}}
\put(2378,-4451){\makebox(0,0)[lb]{\smash{\SetFigFont{9}{10.8}{rm}c}}}
\put(602,-1007){\makebox(0,0)[lb]{\smash{\SetFigFont{12}{14.4}{rm}A}}}
\end{picture}

%% file: paper27.fig9ab.pstex_t
\begin{picture}(0,0)%
\epsfig{file=paper27.fig9ab.pstex}%
\end{picture}%
\setlength{\unitlength}{0.00066700in}%
\begingroup\makeatletter\ifx\SetFigFont\undefined
\def\x#1#2#3#4#5#6#7\relax{\def\x{#1#2#3#4#5#6}}%
\expandafter\x\fmtname xxxxxx\relax \def\y{splain}%
\ifx\x\y   
\gdef\SetFigFont#1#2#3{%
  \ifnum #1<17\tiny\else \ifnum #1<20\small\else
  \ifnum #1<24\normalsize\else \ifnum #1<29\large\else
  \ifnum #1<34\Large\else \ifnum #1<41\LARGE\else
     \huge\fi\fi\fi\fi\fi\fi
  \csname #3\endcsname}%
\else
\gdef\SetFigFont#1#2#3{\begingroup
  \count@#1\relax \ifnum 25<\count@\count@25\fi
  \def\x{\endgroup\@setsize\SetFigFont{#2pt}}%
  \expandafter\x
    \csname \romannumeral\the\count@ pt\expandafter\endcsname
    \csname @\romannumeral\the\count@ pt\endcsname
  \csname #3\endcsname}%
\fi
\fi\endgroup
\begin{picture}(8149,3624)(605,-4139)
\put(2050,-4112){\makebox(0,0)[lb]{\smash{\SetFigFont{10}{12.0}{rm}a}}}
\put(1981,-1096){\makebox(0,0)[lb]{\smash{\SetFigFont{10}{12.0}{rm}$\beta$}}}
\put(7428,-3156){\makebox(0,0)[lb]{\smash{\SetFigFont{10}{12.0}{rm}$n''$}}}
\put(5636,-3156){\makebox(0,0)[lb]{\smash{\SetFigFont{10}{12.0}{rm}$n'$}}}
\put(5994,-1423){\makebox(0,0)[lb]{\smash{\SetFigFont{10}{12.0}{rm}$\tau_1$}}}
\put(7846,-3753){\makebox(0,0)[lb]{\smash{\SetFigFont{10}{12.0}{rm}$p''$}}}
\put(6233,-3753){\makebox(0,0)[lb]{\smash{\SetFigFont{10}{12.0}{rm}$p'$}}}
\put(7069,-4112){\makebox(0,0)[lb]{\smash{\SetFigFont{10}{12.0}{rm}b}}}
\end{picture}

%% file: paper27.fig9cd.pstex_t
\begin{picture}(0,0)%
\epsfig{file=paper27.fig9cd.pstex}%
\end{picture}%
\setlength{\unitlength}{0.00066700in}%
\begingroup\makeatletter\ifx\SetFigFont\undefined
\def\x#1#2#3#4#5#6#7\relax{\def\x{#1#2#3#4#5#6}}%
\expandafter\x\fmtname xxxxxx\relax \def\y{splain}%
\ifx\x\y   
\gdef\SetFigFont#1#2#3{%
  \ifnum #1<17\tiny\else \ifnum #1<20\small\else
  \ifnum #1<24\normalsize\else \ifnum #1<29\large\else
  \ifnum #1<34\Large\else \ifnum #1<41\LARGE\else
     \huge\fi\fi\fi\fi\fi\fi
  \csname #3\endcsname}%
\else
\gdef\SetFigFont#1#2#3{\begingroup
  \count@#1\relax \ifnum 25<\count@\count@25\fi
  \def\x{\endgroup\@setsize\SetFigFont{#2pt}}%
  \expandafter\x
    \csname \romannumeral\the\count@ pt\expandafter\endcsname
    \csname @\romannumeral\the\count@ pt\endcsname
  \csname #3\endcsname}%
\fi
\fi\endgroup
\begin{picture}(8338,4207)(544,-5353)
\put(6902,-2060){\makebox(0,0)[lb]{\smash{\SetFigFont{10}{12.0}{rm}$k''$}}}
\put(7012,-1303){\makebox(0,0)[lb]{\smash{\SetFigFont{10}{12.0}{rm}$q_0$}}}
\put(7159,-5326){\makebox(0,0)[lb]{\smash{\SetFigFont{10}{12.0}{rm}d}}}
\put(2267,-5326){\makebox(0,0)[lb]{\smash{\SetFigFont{10}{12.0}{rm}c}}}
\end{picture}

%% file: paper27.fig10a.pstex_t
\begin{picture}(0,0)%
\epsfig{file=paper27.fig10a.pstex}%
\end{picture}%
\setlength{\unitlength}{0.00066700in}%
\begingroup\makeatletter\ifx\SetFigFont\undefined
\def\x#1#2#3#4#5#6#7\relax{\def\x{#1#2#3#4#5#6}}%
\expandafter\x\fmtname xxxxxx\relax \def\y{splain}%
\ifx\x\y   
\gdef\SetFigFont#1#2#3{%
  \ifnum #1<17\tiny\else \ifnum #1<20\small\else
  \ifnum #1<24\normalsize\else \ifnum #1<29\large\else
  \ifnum #1<34\Large\else \ifnum #1<41\LARGE\else
     \huge\fi\fi\fi\fi\fi\fi
  \csname #3\endcsname}%
\else
\gdef\SetFigFont#1#2#3{\begingroup
  \count@#1\relax \ifnum 25<\count@\count@25\fi
  \def\x{\endgroup\@setsize\SetFigFont{#2pt}}%
  \expandafter\x
    \csname \romannumeral\the\count@ pt\expandafter\endcsname
    \csname @\romannumeral\the\count@ pt\endcsname
  \csname #3\endcsname}%
\fi
\fi\endgroup
\begin{picture}(7750,4206)(634,-4981)
\put(4492,-3886){\makebox(0,0)[lb]{\smash{\SetFigFont{10}{12.0}{rm}a}}}
\put(4192,-1121){\makebox(0,0)[lb]{\smash{\SetFigFont{10}{12.0}{rm}$q_s$}}}
\put(1247,-1482){\makebox(0,0)[lb]{\smash{\SetFigFont{10}{12.0}{rm}k}}}
\put(5251,-4936){\makebox(0,0)[lb]{\smash{\SetFigFont{12}{14.4}{rm}\vspace{0.5cm}}}}
\end{picture}

%% file: paper27.fig10b.pstex_t
\begin{picture}(0,0)%
\epsfig{file=paper27.fig10b.pstex}%
\end{picture}%
\setlength{\unitlength}{0.00066700in}%
\begingroup\makeatletter\ifx\SetFigFont\undefined
\def\x#1#2#3#4#5#6#7\relax{\def\x{#1#2#3#4#5#6}}%
\expandafter\x\fmtname xxxxxx\relax \def\y{splain}%
\ifx\x\y   
\gdef\SetFigFont#1#2#3{%
  \ifnum #1<17\tiny\else \ifnum #1<20\small\else
  \ifnum #1<24\normalsize\else \ifnum #1<29\large\else
  \ifnum #1<34\Large\else \ifnum #1<41\LARGE\else
     \huge\fi\fi\fi\fi\fi\fi
  \csname #3\endcsname}%
\else
\gdef\SetFigFont#1#2#3{\begingroup
  \count@#1\relax \ifnum 25<\count@\count@25\fi
  \def\x{\endgroup\@setsize\SetFigFont{#2pt}}%
  \expandafter\x
    \csname \romannumeral\the\count@ pt\expandafter\endcsname
    \csname @\romannumeral\the\count@ pt\endcsname
  \csname #3\endcsname}%
\fi
\fi\endgroup
\begin{picture}(7838,2826)(590,-3106)
\put(3751,-436){\makebox(0,0)[lb]{\smash{\SetFigFont{10}{12.0}{rm}$q_s\,q'_s$}}}
\put(3901,-1036){\makebox(0,0)[lb]{\smash{\SetFigFont{10}{12.0}{rm}q"}}}
\put(1456,-1044){\makebox(0,0)[lb]{\smash{\SetFigFont{10}{12.0}{rm}$V_3$}}}
\put(4509,-2326){\makebox(0,0)[lb]{\smash{\SetFigFont{10}{12.0}{rm}b}}}
\put(5176,-3061){\makebox(0,0)[lb]{\smash{\SetFigFont{12}{14.4}{rm}\vspace{0.5cm}}}}
\put(3520,-763){\makebox(0,0)[lb]{\smash{\SetFigFont{8}{9.6}{rm}1}}}
\put(4386,-763){\makebox(0,0)[lb]{\smash{\SetFigFont{8}{9.6}{rm}2}}}
\end{picture}

%% file: paper27.fig10c.pstex_t
\begin{picture}(0,0)%
\epsfig{file=paper27.fig10c.pstex}%
\end{picture}%
\setlength{\unitlength}{0.00066700in}%
\begingroup\makeatletter\ifx\SetFigFont\undefined
\def\x#1#2#3#4#5#6#7\relax{\def\x{#1#2#3#4#5#6}}%
\expandafter\x\fmtname xxxxxx\relax \def\y{splain}%
\ifx\x\y   
\gdef\SetFigFont#1#2#3{%
  \ifnum #1<17\tiny\else \ifnum #1<20\small\else
  \ifnum #1<24\normalsize\else \ifnum #1<29\large\else
  \ifnum #1<34\Large\else \ifnum #1<41\LARGE\else
     \huge\fi\fi\fi\fi\fi\fi
  \csname #3\endcsname}%
\else
\gdef\SetFigFont#1#2#3{\begingroup
  \count@#1\relax \ifnum 25<\count@\count@25\fi
  \def\x{\endgroup\@setsize\SetFigFont{#2pt}}%
  \expandafter\x
    \csname \romannumeral\the\count@ pt\expandafter\endcsname
    \csname @\romannumeral\the\count@ pt\endcsname
  \csname #3\endcsname}%
\fi
\fi\endgroup
\begin{picture}(7753,2106)(590,-2795)
\put(4467,-2776){\makebox(0,0)[lb]{\smash{\SetFigFont{10}{12.0}{rm}c}}}
\put(1508,-1629){\makebox(0,0)[lb]{\smash{\SetFigFont{10}{12.0}{rm}k}}}
\put(1689,-844){\makebox(0,0)[lb]{\smash{\SetFigFont{10}{12.0}{rm}$q_s$}}}
\put(1206,-1267){\makebox(0,0)[lb]{\smash{\SetFigFont{10}{12.0}{rm}$V_3$}}}
\end{picture}

%% file: paper27.fig11.pstex_t
\begin{picture}(0,0)%
\epsfig{file=paper27.fig11.pstex}%
\end{picture}%
\setlength{\unitlength}{0.00070000in}%
\begingroup\makeatletter\ifx\SetFigFont\undefined
\def\x#1#2#3#4#5#6#7\relax{\def\x{#1#2#3#4#5#6}}%
\expandafter\x\fmtname xxxxxx\relax \def\y{splain}%
\ifx\x\y   
\gdef\SetFigFont#1#2#3{%
  \ifnum #1<17\tiny\else \ifnum #1<20\small\else
  \ifnum #1<24\normalsize\else \ifnum #1<29\large\else
  \ifnum #1<34\Large\else \ifnum #1<41\LARGE\else
     \huge\fi\fi\fi\fi\fi\fi
  \csname #3\endcsname}%
\else
\gdef\SetFigFont#1#2#3{\begingroup
  \count@#1\relax \ifnum 25<\count@\count@25\fi
  \def\x{\endgroup\@setsize\SetFigFont{#2pt}}%
  \expandafter\x
    \csname \romannumeral\the\count@ pt\expandafter\endcsname
    \csname @\romannumeral\the\count@ pt\endcsname
  \csname #3\endcsname}%
\fi
\fi\endgroup
\begin{picture}(8134,2583)(857,-2743)
\put(857,-666){\makebox(0,0)[lb]{\smash{\SetFigFont{9}{10.8}{rm}k}}}
\put(1272,-471){\makebox(0,0)[lb]{\smash{\SetFigFont{9}{10.8}{rm}$\nu$}}}
\put(1304,-1446){\makebox(0,0)[lb]{\smash{\SetFigFont{9}{10.8}{rm}$\mu$}}}
\put(1655,-722){\makebox(0,0)[lb]{\smash{\SetFigFont{9}{10.8}{rm}$\sigma$}}}
\put(7600,-965){\makebox(0,0)[lb]{\smash{\SetFigFont{6}{7.2}{rm}$\sigma$}}}
\put(1711,-2086){\makebox(0,0)[lb]{\smash{\SetFigFont{8}{9.6}{rm}a}}}
\put(5176,-2086){\makebox(0,0)[lb]{\smash{\SetFigFont{8}{9.6}{rm}b}}}
\put(8011,-2086){\makebox(0,0)[lb]{\smash{\SetFigFont{8}{9.6}{rm}c}}}
\put(4141,-2716){\makebox(0,0)[lb]{\smash{\SetFigFont{8}{9.6}{rm}\vspace{0.5cm}}}}
\put(7494,-1478){\makebox(0,0)[lb]{\smash{\SetFigFont{6}{7.2}{rm}q}}}
\put(1527,-1112){\makebox(0,0)[lb]{\smash{\SetFigFont{9}{10.8}{rm}q}}}
\put(4153,-295){\makebox(0,0)[lb]{\smash{\SetFigFont{9}{10.8}{rm}$q_i$}}}
\put(4153,-702){\makebox(0,0)[lb]{\smash{\SetFigFont{9}{10.8}{rm}$\sigma'$}}}
\put(4878,-702){\makebox(0,0)[lb]{\smash{\SetFigFont{9}{10.8}{rm}$\sigma$}}}
\put(4658,-1206){\makebox(0,0)[lb]{\smash{\SetFigFont{9}{10.8}{rm}$\mu$}}}
\end{picture}

%% file: paper27.fig12.pstex_t
\begin{picture}(0,0)%
\epsfig{file=paper27.fig12.pstex}%
\end{picture}%
\setlength{\unitlength}{0.00070000in}%
\begingroup\makeatletter\ifx\SetFigFont\undefined
\def\x#1#2#3#4#5#6#7\relax{\def\x{#1#2#3#4#5#6}}%
\expandafter\x\fmtname xxxxxx\relax \def\y{splain}%
\ifx\x\y   
\gdef\SetFigFont#1#2#3{%
  \ifnum #1<17\tiny\else \ifnum #1<20\small\else
  \ifnum #1<24\normalsize\else \ifnum #1<29\large\else
  \ifnum #1<34\Large\else \ifnum #1<41\LARGE\else
     \huge\fi\fi\fi\fi\fi\fi
  \csname #3\endcsname}%
\else
\gdef\SetFigFont#1#2#3{\begingroup
  \count@#1\relax \ifnum 25<\count@\count@25\fi
  \def\x{\endgroup\@setsize\SetFigFont{#2pt}}%
  \expandafter\x
    \csname \romannumeral\the\count@ pt\expandafter\endcsname
    \csname @\romannumeral\the\count@ pt\endcsname
  \csname #3\endcsname}%
\fi
\fi\endgroup
\begin{picture}(4524,3489)(2464,-3091)
\put(4501,-601){\makebox(0,0)[lb]{\smash{\SetFigFont{17}{20.4}{rm}N}}}
\put(4501,-2941){\makebox(0,0)[lb]{\smash{\SetFigFont{17}{20.4}{rm}N}}}
\put(3061,254){\makebox(0,0)[lb]{\smash{\SetFigFont{10}{12.0}{rm}\vspace{0.5cm}}}}
\end{picture}

%% file: paper27.fig13ab.pstex_t
\begin{picture}(0,0)%
\epsfig{file=paper27.fig13ab.pstex}%
\end{picture}%
\setlength{\unitlength}{0.00066700in}%
\begingroup\makeatletter\ifx\SetFigFont\undefined
\def\x#1#2#3#4#5#6#7\relax{\def\x{#1#2#3#4#5#6}}%
\expandafter\x\fmtname xxxxxx\relax \def\y{splain}%
\ifx\x\y   
\gdef\SetFigFont#1#2#3{%
  \ifnum #1<17\tiny\else \ifnum #1<20\small\else
  \ifnum #1<24\normalsize\else \ifnum #1<29\large\else
  \ifnum #1<34\Large\else \ifnum #1<41\LARGE\else
     \huge\fi\fi\fi\fi\fi\fi
  \csname #3\endcsname}%
\else
\gdef\SetFigFont#1#2#3{\begingroup
  \count@#1\relax \ifnum 25<\count@\count@25\fi
  \def\x{\endgroup\@setsize\SetFigFont{#2pt}}%
  \expandafter\x
    \csname \romannumeral\the\count@ pt\expandafter\endcsname
    \csname @\romannumeral\the\count@ pt\endcsname
  \csname #3\endcsname}%
\fi
\fi\endgroup
\begin{picture}(7968,4651)(861,-5280)
\put(6927,-5253){\makebox(0,0)[lb]{\smash{\SetFigFont{10}{12.0}{rm}b}}}
\put(4383,-761){\makebox(0,0)[lb]{\smash{\SetFigFont{12}{14.4}{rm}2}}}
\put(1163,-856){\makebox(0,0)[lb]{\smash{\SetFigFont{11}{13.2}{rm}h}}}
\put(1163,-3827){\makebox(0,0)[lb]{\smash{\SetFigFont{11}{13.2}{rm}d}}}
\put(2739,-5253){\makebox(0,0)[lb]{\smash{\SetFigFont{10}{12.0}{rm}a}}}
\end{picture}

%% file: paper27.fig13c.pstex_t
\begin{picture}(0,0)%
\epsfig{file=paper27.fig13c.pstex}%
\end{picture}%
\setlength{\unitlength}{0.00066700in}%
\begingroup\makeatletter\ifx\SetFigFont\undefined
\def\x#1#2#3#4#5#6#7\relax{\def\x{#1#2#3#4#5#6}}%
\expandafter\x\fmtname xxxxxx\relax \def\y{splain}%
\ifx\x\y   
\gdef\SetFigFont#1#2#3{%
  \ifnum #1<17\tiny\else \ifnum #1<20\small\else
  \ifnum #1<24\normalsize\else \ifnum #1<29\large\else
  \ifnum #1<34\Large\else \ifnum #1<41\LARGE\else
     \huge\fi\fi\fi\fi\fi\fi
  \csname #3\endcsname}%
\else
\gdef\SetFigFont#1#2#3{\begingroup
  \count@#1\relax \ifnum 25<\count@\count@25\fi
  \def\x{\endgroup\@setsize\SetFigFont{#2pt}}%
  \expandafter\x
    \csname \romannumeral\the\count@ pt\expandafter\endcsname
    \csname @\romannumeral\the\count@ pt\endcsname
  \csname #3\endcsname}%
\fi
\fi\endgroup
\begin{picture}(8020,4009)(889,-5410)
\put(8909,-1533){\makebox(0,0)[lb]{\smash{\SetFigFont{12}{14.4}{rm}2}}}
\put(4651,-5386){\makebox(0,0)[lb]{\smash{\SetFigFont{12}{14.4}{rm}c}}}
\end{picture}

%% file: paper27.fig14a.pstex_t
\begin{picture}(0,0)%
\epsfig{file=paper27.fig14a.pstex}%
\end{picture}%
\setlength{\unitlength}{0.00070000in}%
\begingroup\makeatletter\ifx\SetFigFont\undefined
\def\x#1#2#3#4#5#6#7\relax{\def\x{#1#2#3#4#5#6}}%
\expandafter\x\fmtname xxxxxx\relax \def\y{splain}%
\ifx\x\y   
\gdef\SetFigFont#1#2#3{%
  \ifnum #1<17\tiny\else \ifnum #1<20\small\else
  \ifnum #1<24\normalsize\else \ifnum #1<29\large\else
  \ifnum #1<34\Large\else \ifnum #1<41\LARGE\else
     \huge\fi\fi\fi\fi\fi\fi
  \csname #3\endcsname}%
\else
\gdef\SetFigFont#1#2#3{\begingroup
  \count@#1\relax \ifnum 25<\count@\count@25\fi
  \def\x{\endgroup\@setsize\SetFigFont{#2pt}}%
  \expandafter\x
    \csname \romannumeral\the\count@ pt\expandafter\endcsname
    \csname @\romannumeral\the\count@ pt\endcsname
  \csname #3\endcsname}%
\fi
\fi\endgroup
\begin{picture}(7959,4277)(947,-4284)
\put(7200,-3102){\makebox(0,0)[lb]{\smash{\SetFigFont{14}{16.8}{rm}N}}}
\put(7388,-3177){\makebox(0,0)[lb]{\smash{\SetFigFont{14}{16.8}{rm}4}}}
\put(7200,-279){\makebox(0,0)[lb]{\smash{\SetFigFont{14}{16.8}{rm}N}}}
\put(7388,-316){\makebox(0,0)[lb]{\smash{\SetFigFont{14}{16.8}{rm}4}}}
\put(1420,-1688){\makebox(0,0)[lb]{\smash{\SetFigFont{9}{10.8}{rm}$\omega$}}}
\put(947,-1616){\makebox(0,0)[lb]{\smash{\SetFigFont{14}{16.8}{rm}disc }}}
\put(4544,-4254){\makebox(0,0)[lb]{\smash{\SetFigFont{9}{10.8}{rm}\vspace{0.5cm}}}}
\put(5143,-3779){\makebox(0,0)[lb]{\smash{\SetFigFont{10}{12.0}{rm}a}}}
\put(7802,-1257){\makebox(0,0)[lb]{\smash{\SetFigFont{8}{9.6}{rm}4}}}
\put(7351,-1257){\makebox(0,0)[lb]{\smash{\SetFigFont{8}{9.6}{rm}3}}}
\put(6899,-1257){\makebox(0,0)[lb]{\smash{\SetFigFont{8}{9.6}{rm}2}}}
\put(6410,-1257){\makebox(0,0)[lb]{\smash{\SetFigFont{8}{9.6}{rm}1}}}
\end{picture}

%% file: paper27.fig14b.pstex_t
\begin{picture}(0,0)%
\epsfig{file=paper27.fig14b.pstex}%
\end{picture}%
\setlength{\unitlength}{0.00065600in}%
\begingroup\makeatletter\ifx\SetFigFont\undefined
\def\x#1#2#3#4#5#6#7\relax{\def\x{#1#2#3#4#5#6}}%
\expandafter\x\fmtname xxxxxx\relax \def\y{splain}%
\ifx\x\y   
\gdef\SetFigFont#1#2#3{%
  \ifnum #1<17\tiny\else \ifnum #1<20\small\else
  \ifnum #1<24\normalsize\else \ifnum #1<29\large\else
  \ifnum #1<34\Large\else \ifnum #1<41\LARGE\else
     \huge\fi\fi\fi\fi\fi\fi
  \csname #3\endcsname}%
\else
\gdef\SetFigFont#1#2#3{\begingroup
  \count@#1\relax \ifnum 25<\count@\count@25\fi
  \def\x{\endgroup\@setsize\SetFigFont{#2pt}}%
  \expandafter\x
    \csname \romannumeral\the\count@ pt\expandafter\endcsname
    \csname @\romannumeral\the\count@ pt\endcsname
  \csname #3\endcsname}%
\fi
\fi\endgroup
\begin{picture}(9481,7157)(837,-7080)
\put(5088,-5132){\makebox(0,0)[lb]{\smash{\SetFigFont{14}{16.8}{rm}N}}}
\put(4667,-3601){\makebox(0,0)[lb]{\smash{\SetFigFont{20}{24.0}{rm}+}}}
\put(7840,-1176){\makebox(0,0)[lb]{\smash{\SetFigFont{34}{40.8}{rm}=}}}
\put(5132,-394){\makebox(0,0)[lb]{\smash{\SetFigFont{14}{16.8}{rm}4}}}
\put(4878,-285){\makebox(0,0)[lb]{\smash{\SetFigFont{14}{16.8}{rm}N}}}
\put(5935,-1889){\makebox(0,0)[lb]{\smash{\SetFigFont{9}{10.8}{rm}4}}}
\put(5303,-5204){\makebox(0,0)[lb]{\smash{\SetFigFont{12}{14.4}{rm}4}}}
\put(4951,-6586){\makebox(0,0)[lb]{\smash{\SetFigFont{11}{13.2}{rm}b}}}
\put(4286,-7041){\makebox(0,0)[lb]{\smash{\SetFigFont{11}{13.2}{rm}\vspace{0.5cm}}}}
\put(2471,-5591){\makebox(0,0)[lb]{\smash{\SetFigFont{20}{24.0}{rm}+}}}
\put(7672,-394){\makebox(0,0)[lb]{\smash{\SetFigFont{20}{24.0}{rm}$\gamma^*$}}}
\put(1536,-394){\makebox(0,0)[lb]{\smash{\SetFigFont{20}{24.0}{rm}$\gamma^*$}}}
\put(5427,-4991){\makebox(0,0)[lb]{\smash{\SetFigFont{12}{14.4}{rm}I}}}
\put(5303,-1889){\makebox(0,0)[lb]{\smash{\SetFigFont{9}{10.8}{rm}3}}}
\put(4667,-1889){\makebox(0,0)[lb]{\smash{\SetFigFont{9}{10.8}{rm}2}}}
\put(4030,-1889){\makebox(0,0)[lb]{\smash{\SetFigFont{9}{10.8}{rm}1}}}
\put(7840,-1176){\makebox(0,0)[lb]{\smash{\SetFigFont{34}{40.8}{rm}=}}}
\put(5132,-394){\makebox(0,0)[lb]{\smash{\SetFigFont{14}{16.8}{rm}4}}}
\put(4878,-285){\makebox(0,0)[lb]{\smash{\SetFigFont{14}{16.8}{rm}N}}}
\put(5935,-1889){\makebox(0,0)[lb]{\smash{\SetFigFont{9}{10.8}{rm}4}}}
\put(5303,-1889){\makebox(0,0)[lb]{\smash{\SetFigFont{9}{10.8}{rm}3}}}
\put(4667,-1889){\makebox(0,0)[lb]{\smash{\SetFigFont{9}{10.8}{rm}2}}}
\put(4667,-3601){\makebox(0,0)[lb]{\smash{\SetFigFont{20}{24.0}{rm}+}}}
\put(7672,-394){\makebox(0,0)[lb]{\smash{\SetFigFont{20}{24.0}{rm}$\gamma^*$}}}
\put(1536,-394){\makebox(0,0)[lb]{\smash{\SetFigFont{20}{24.0}{rm}$\gamma^*$}}}
\put(5427,-4991){\makebox(0,0)[lb]{\smash{\SetFigFont{12}{14.4}{rm}I}}}
\put(5303,-5204){\makebox(0,0)[lb]{\smash{\SetFigFont{12}{14.4}{rm}4}}}
\put(5088,-5132){\makebox(0,0)[lb]{\smash{\SetFigFont{14}{16.8}{rm}N}}}
\put(4030,-1889){\makebox(0,0)[lb]{\smash{\SetFigFont{9}{10.8}{rm}1}}}
\end{picture}

%% file: paper27.fig14c.pstex_t
\begin{picture}(0,0)%
\epsfig{file=paper27.fig14c.pstex}%
\end{picture}%
\setlength{\unitlength}{0.00065600in}%
\begingroup\makeatletter\ifx\SetFigFont\undefined
\def\x#1#2#3#4#5#6#7\relax{\def\x{#1#2#3#4#5#6}}%
\expandafter\x\fmtname xxxxxx\relax \def\y{splain}%
\ifx\x\y   
\gdef\SetFigFont#1#2#3{%
  \ifnum #1<17\tiny\else \ifnum #1<20\small\else
  \ifnum #1<24\normalsize\else \ifnum #1<29\large\else
  \ifnum #1<34\Large\else \ifnum #1<41\LARGE\else
     \huge\fi\fi\fi\fi\fi\fi
  \csname #3\endcsname}%
\else
\gdef\SetFigFont#1#2#3{\begingroup
  \count@#1\relax \ifnum 25<\count@\count@25\fi
  \def\x{\endgroup\@setsize\SetFigFont{#2pt}}%
  \expandafter\x
    \csname \romannumeral\the\count@ pt\expandafter\endcsname
    \csname @\romannumeral\the\count@ pt\endcsname
  \csname #3\endcsname}%
\fi
\fi\endgroup
\begin{picture}(8022,4690)(968,-5043)
\put(4897,-5025){\makebox(0,0)[lb]{\smash{\SetFigFont{9}{10.8}{rm}c}}}
\put(4337,-2351){\makebox(0,0)[lb]{\smash{\SetFigFont{29}{34.8}{rm}=}}}
\put(2533,-2164){\makebox(0,0)[lb]{\smash{\SetFigFont{9}{10.8}{rm}I}}}
\put(2409,-2475){\makebox(0,0)[lb]{\smash{\SetFigFont{9}{10.8}{rm}4}}}
\put(2222,-2351){\makebox(0,0)[lb]{\smash{\SetFigFont{12}{14.4}{rm}N}}}
\put(4897,-5025){\makebox(0,0)[lb]{\smash{\SetFigFont{9}{10.8}{rm}c}}}
\put(4337,-2351){\makebox(0,0)[lb]{\smash{\SetFigFont{29}{34.8}{rm}=}}}
\put(2533,-2164){\makebox(0,0)[lb]{\smash{\SetFigFont{9}{10.8}{rm}I}}}
\put(2409,-2475){\makebox(0,0)[lb]{\smash{\SetFigFont{9}{10.8}{rm}4}}}
\put(2222,-2351){\makebox(0,0)[lb]{\smash{\SetFigFont{12}{14.4}{rm}N}}}
\put(2222,-2351){\makebox(0,0)[lb]{\smash{\SetFigFont{12}{14.4}{rm}N}}}
\put(2409,-2475){\makebox(0,0)[lb]{\smash{\SetFigFont{9}{10.8}{rm}4}}}
\put(2533,-2164){\makebox(0,0)[lb]{\smash{\SetFigFont{9}{10.8}{rm}I}}}
\put(4337,-2351){\makebox(0,0)[lb]{\smash{\SetFigFont{29}{34.8}{rm}=}}}
\put(4897,-5025){\makebox(0,0)[lb]{\smash{\SetFigFont{9}{10.8}{rm}c}}}
\end{picture}

%% file: paper27.fig15ab.pstex_t
\begin{picture}(0,0)%
\epsfig{file=paper27.fig15ab.pstex}%
\end{picture}%
\setlength{\unitlength}{0.00066700in}%
\begingroup\makeatletter\ifx\SetFigFont\undefined
\def\x#1#2#3#4#5#6#7\relax{\def\x{#1#2#3#4#5#6}}%
\expandafter\x\fmtname xxxxxx\relax \def\y{splain}%
\ifx\x\y   
\gdef\SetFigFont#1#2#3{%
  \ifnum #1<17\tiny\else \ifnum #1<20\small\else
  \ifnum #1<24\normalsize\else \ifnum #1<29\large\else
  \ifnum #1<34\Large\else \ifnum #1<41\LARGE\else
     \huge\fi\fi\fi\fi\fi\fi
  \csname #3\endcsname}%
\else
\gdef\SetFigFont#1#2#3{\begingroup
  \count@#1\relax \ifnum 25<\count@\count@25\fi
  \def\x{\endgroup\@setsize\SetFigFont{#2pt}}%
  \expandafter\x
    \csname \romannumeral\the\count@ pt\expandafter\endcsname
    \csname @\romannumeral\the\count@ pt\endcsname
  \csname #3\endcsname}%
\fi
\fi\endgroup
\begin{picture}(8080,5285)(683,-5047)
\put(4651,-5011){\makebox(0,0)[lb]{\smash{\SetFigFont{10}{12.0}{rm}\vspace{0.5cm}}}}
\put(6839,-4570){\makebox(0,0)[lb]{\smash{\SetFigFont{10}{12.0}{rm}b}}}
\put(2497,-4570){\makebox(0,0)[lb]{\smash{\SetFigFont{10}{12.0}{rm}a}}}
\end{picture}

%% file: paper27.fig15c.pstex_t
\begin{picture}(0,0)%
\epsfig{file=paper27.fig15c.pstex}%
\end{picture}%
\setlength{\unitlength}{0.00066700in}%
\begingroup\makeatletter\ifx\SetFigFont\undefined
\def\x#1#2#3#4#5#6#7\relax{\def\x{#1#2#3#4#5#6}}%
\expandafter\x\fmtname xxxxxx\relax \def\y{splain}%
\ifx\x\y   
\gdef\SetFigFont#1#2#3{%
  \ifnum #1<17\tiny\else \ifnum #1<20\small\else
  \ifnum #1<24\normalsize\else \ifnum #1<29\large\else
  \ifnum #1<34\Large\else \ifnum #1<41\LARGE\else
     \huge\fi\fi\fi\fi\fi\fi
  \csname #3\endcsname}%
\else
\gdef\SetFigFont#1#2#3{\begingroup
  \count@#1\relax \ifnum 25<\count@\count@25\fi
  \def\x{\endgroup\@setsize\SetFigFont{#2pt}}%
  \expandafter\x
    \csname \romannumeral\the\count@ pt\expandafter\endcsname
    \csname @\romannumeral\the\count@ pt\endcsname
  \csname #3\endcsname}%
\fi
\fi\endgroup
\begin{picture}(3879,4827)(2639,-4726)
\put(4576,-4711){\makebox(0,0)[lb]{\smash{\SetFigFont{7}{8.4}{rm}c}}}
\end{picture}

%% file: paper27.fig16.pstex_t
\begin{picture}(0,0)%
\epsfig{file=paper27.fig16.pstex}%
\end{picture}%
\setlength{\unitlength}{0.00070000in}%
\begingroup\makeatletter\ifx\SetFigFont\undefined
\def\x#1#2#3#4#5#6#7\relax{\def\x{#1#2#3#4#5#6}}%
\expandafter\x\fmtname xxxxxx\relax \def\y{splain}%
\ifx\x\y   
\gdef\SetFigFont#1#2#3{%
  \ifnum #1<17\tiny\else \ifnum #1<20\small\else
  \ifnum #1<24\normalsize\else \ifnum #1<29\large\else
  \ifnum #1<34\Large\else \ifnum #1<41\LARGE\else
     \huge\fi\fi\fi\fi\fi\fi
  \csname #3\endcsname}%
\else
\gdef\SetFigFont#1#2#3{\begingroup
  \count@#1\relax \ifnum 25<\count@\count@25\fi
  \def\x{\endgroup\@setsize\SetFigFont{#2pt}}%
  \expandafter\x
    \csname \romannumeral\the\count@ pt\expandafter\endcsname
    \csname @\romannumeral\the\count@ pt\endcsname
  \csname #3\endcsname}%
\fi
\fi\endgroup
\begin{picture}(8056,5325)(848,-5044)
\put(2705,-5012){\makebox(0,0)[lb]{\smash{\SetFigFont{12}{14.4}{rm}a}}}
\put(6887,-5012){\makebox(0,0)[lb]{\smash{\SetFigFont{12}{14.4}{rm}b}}}
\end{picture}